\documentclass[a4paper]{spie}

\usepackage{graphicx, subfigure}
\usepackage{float}
\usepackage{epsfig}

\title{Profile reconstruction of grazing-incidence X-ray mirrors from intra-focal X-ray full imaging}

\author{D. Spiga\supit{1}, S. Basso\supit{1}, M. Bavdaz\supit{2}, V. Burwitz\supit{3}, M. Civitani\supit{1}, O. Citterio\supit{1}, M. Ghigo\supit{1},  G.~Hartner\supit{3}, B. Menz\supit{3}, G. Pareschi\supit{1}, L. Proserpio\supit{3}, B. Salmaso\supit{1,4}, G. Tagliaferri\supit{1}, E. Wille\supit{2}
\skiplinehalf
\supit{1}INAF/ Osservatorio Astronomico di Brera, Via E. Bianchi 46, 23807 Merate, Italy\\
\supit{2}European Space Agency, ESTEC, Noordwjik, Netherlands\\
\supit{3}Max-Planck Instit\"ut fur Extraterrestr. Physik, Giessenbach Stra$\beta$e, 85471 Garching, Germany\\
\supit{4}Universit\`a dell$'$Insubria, Via Valleggio 11, 22100 Como, Italy}

\authorinfo{}

\begin{document}
\maketitle 
\begin{abstract}
The optics of a number of future X-ray telescopes will have very long focal lengths (10 -- 20 m), and will consist of a number of nested/stacked thin, grazing-incidence mirrors. The optical quality characterization of a real mirror can be obtained via profile metrology, and the Point Spread Function of the mirror can be derived via one of the standard computation methods. However, in practical cases it can be difficult to access the optical surfaces of densely stacked mirror shells, after they have been assembled, using the widespread metrological tools. For this reason, the assessment of the imaging resolution of a system of mirrors is better obtained via a direct, full-illumination test in X-rays. If the focus cannot be reached, an intra-focus test can be performed, and the image can be compared with the simulation results based on the metrology, if available. However, until today no quantitative information was extracted from a full-illumination, intra-focal exposure. In this work we show that, if the detector is located at an optimal distance from the mirror, the intensity variations of the intra-focal, full-illumination image in single reflection can be used to reconstruct the profile of the mirror surface, without the need of a wavefront sensor. The Point Spread Function can be subsequently computed from the reconstructed mirror shape. We show the application of this method to an intra-focal (8 m distance from mirror) test performed at PANTER on an optical module prototype made of hot-slumped glass foils with a 20 m focal length, from which we could derive an expected imaging quality near 16 arcsec HEW.
\end{abstract}


\section{Introduction}
\label{intro}
The optics of future X-ray imaging telescopes will need to conjugate high angular resolutions and high effective areas, with a low mass/effective area ratio in order to ensure the telescope operation in space. For this reason, the optical module will comprise thin, densely stacked, precisely figured mirrors made of a lightweight material like, e.g., Silicon or glass. The shallow incidence angles ($<$ 1 deg) and the large mirror apertures ($>$ 1 m) at play entail focal lengths of tenths of meters and make impossible to manufacture monolithic mirrors to be nested in modules. Rather, a modular approach has to be adopted, assembling stacks of smaller, tightly stacked foil segments after endowing them with a focusing profile like the Wolter's\cite{VanSpey}: the resulting blocks are subsequently assembled and accurately aligned into a supporting structure to return an optical module with the desired properties. Adopting this methodology, two approaches are currently being pursued in Europe, aimed at the development of the optics for the ATHENA (formerly IXO) X-ray telescope\cite{Bavdaz}: one based on Silicon Pore Optics\cite{SPO} and the other based on hot-slumped glasses\cite{CivitaniOE}. In particular, the hot-slumping approach has been developed in the last years at INAF/OAB, in the ESA-led project "IXO backup optics with slumped glasses", aiming at demonstrating the feasibility of high angular resolutions (i.e., $\sim$15 arcsec HEW at this first stage, with a final goal of 5~arcsec) via a proper optimization of the slumping\cite{Proserpio, Ghigo} and the integration\cite{IMA} processes. Even if ATHENA was not selected in response to the L1 call, the results achieved will be very useful in view of the possible ATHENA+ selection for L2/L3 in fall 2013. An independent development of segmented optics based on hot-slumped glasses has been carried out in the USA to manufacture the optical modules of the presently-operated NuSTAR\cite{SlumpingUS} X-ray telescope. 

\begin{figure}[H]
	\centering
	\begin{tabular}{ll}
		\includegraphics[width = 0.46 \textwidth]{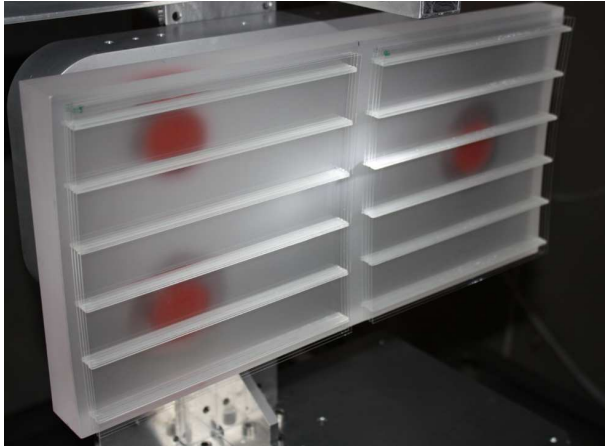}&
		 \includegraphics[width = 0.46 \textwidth]{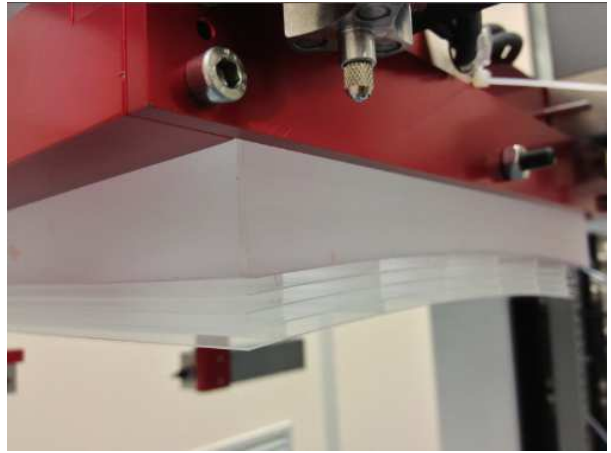}\\
		 \footnotesize A) &  \footnotesize B)
	\end{tabular}
	\caption{The Proof of Concept No. 2 (PoC\#2) optic based on hot-slumped glass foils\cite{Civitani}. A) front view: through the glass foil stack, the stiffening glass ribs and the glass backplane can be seen. The three red circles mark the supporting pads. B) sideways view: the stack structure of the 4 co-focal mirror pairs is clearly visible. The mirror pairs are denoted as PP0, \ldots PP3, from bottom to top in this picture, where the optic is mounted in the Integration MAchine (IMA \cite{IMA}). }
	\label{fig:PoC2}
\end{figure}
\vspace{-2mm}
Regardless of the technique adopted, a fundamental step is the assessment of the X-ray optical quality of the modular elements produced, in terms of Point Spread Function (PSF) or Half-Energy-Width (HEW, i.e., twice the median value of the PSF). Even though this can be effectively simulated from the analysis of the surface profile and roughness measurements\cite{RaiSpi1, RaiSpi2}, a direct proof of the attained angular resolution comes from a direct, in-focus measurement. Moreover, metrology tools can seldom be used to measure the surface of mirrors after the integration, which are often closely spaced in dense stacks. The measurement in X-rays can be done in a large X-ray facility like PANTER\cite{PANTER1, PANTER2} (MPE Neuried, Germany), which in addition was extended\cite{PANTERext} in 2012 to be able to match focal lengths up to 20 m, the one foreseen for the IXO telescope (plus a $\sim$4 m shift caused by the X-ray beam divergence). The PANTER extension was successfully used, for the first time, to measure in focus and in full illumination the XOU\_BB (X-ray Optical Unit BreadBoard\cite{CivitaniOE}, see Tab.~\ref{tab:XOUs}) manufactured at INAF/OAB by hot slumping of thin glass foils. However, the utilization of the extension could not be scheduled in spring 2013 to perform the in-focus tests of the subsequent demonstrator optic, the Proof of Concept No. 2 (PoC\#2, Fig.~\ref{fig:PoC2}). Hence, the PoC\#2 tests at PANTER had to be done in intra-focal position, even if a considerably better optical quality was expected from the PoC\#2, owing to several improvements in the equipment and the techniques adopted in the process. More details on the PoC\#2 realization can be retrieved from another paper of this SPIE volume\cite{Civitani}. In this paper we concentrate on the reconstruction of the mirror shape from the intra-focal exposure measured at PANTER, to consequently derive the PSF at the focal plane.

\begin{table}[hbt]
	\centering
	\begin{tabular}{lllll}
	\hline
	\it Item & \it Date & \it Tested & \it Single shell tested & \it Method used\\
	\hline
	PoC\#1 & Dec 2011 & Intrafocus & HEW $\approx$ 80 arcsec & comp. from surface mapping\\
	XOU\_BB & Aug 2012 & In focus & HEW $\approx$ 60 arcsec & measured in focus \\
	PoC\#2 & Apr 2013 & Intrafocus & HEW $\approx$ 16 arcsec & {\bf this paper}\\
	\hline
	\end{tabular}
	\vspace{1.5mm}
	\caption{The demonstrators manufactured for the "IXO backup optics with slumped glasses" project, with the X-ray tests performed at PANTER at the 0.27~keV X-ray energy, with the methods used to assess the HEW. The derivation of the in-focus PSF for the PP0 shell of the PoC\#2 from the intra-focal test is the subject of this work.}
	\label{tab:XOUs}
\end{table}

A suitable way to derive a mirror optical quality, given an out-of-focus exposure, is the wavefront detection using a Hartmann sensor\cite{Hartmann}. This device implements the well-known Hartmann test and performs a sampling/discretization of the X-ray beam focused by the mirror under test, by means of a "sieve" plate. The beamlet deflection, as measured by a position sensitive detector allows one to reconstruct the sampled wavefront and consequently extrapolate to focus. However, at very shallow angles the Hartmann plate requires extremely small and closely spaced holes in order to sample the longitudinal profile of the mirror. A Hartmann test concept for X-ray telescopes is proposed in another paper of this volume\cite{Saha}.

The X-ray pencil beam is another suitable technique to check the performances of a long focal length optic, and in fact it is routinely used in the SPO performance tests\cite{SPOtest} at the PTB laboratory of the BESSY synchrotron. Succinctly put, this approach requires a thin and intense X-ray beam to probe the optic pore by pore, obtaining a set of partial focal spot images that can be superposed compensating the offset of the optic under test. If the detector cannot be placed at the correct focal distance, the information on the incidence position on the optic and the final ray position on the detector in intra-focal position can be geometrically extrapolated to focus, thereby reconstructing the focal spot.

In this paper we propose {\it an alternative method to derive the profile of a grazing-incidence mirror from an X-ray full-illumination exposure, measured intra-focus}. The experimental setup does not require any particular equipment but the one available at PANTER in its pre-extension configuration (Sect.~\ref{tests}). Unlike often done when using Hartmann sensors, the detector has to be placed quite near to the mirror in order to -- among other things -- record a sufficiently spatially resolved intra-focal image; the mirror profile is then reconstructed from the brightness variations of the image, using equations (Sect.~\ref{recon}) adapted from a beam-shaping formalism proposed recently\cite{Spiga13}. As of today, the analysis is possible only in single reflection setup, hence the mirrors under test have to be measured in single reflection by properly setting the incidence angle (Fig.~\ref{fig:align2}). Because of the obstructions in the stack, the single reflection measurements were possible only for the outermost layer of the stack, named PP0 (Fig.~\ref{fig:PoC2}). The analysis is so far limited to the longitudinal profiles, which in grazing-incidence overwhelmingly dominate the roundness errors in determining the PSF. Once obtained the mirror profile (Sect.~\ref{profile}), one can compute the expected PSF in focus for double reflection (Sect.~\ref{concl}) either by ray-tracing or by one of the known methods to account for the effect of surface roughness\cite{RaiSpi2}. The final result derived {\it is a predicted HEW of 15.5 arcsec at 0.27~keV} for the outermost mirror pair of the PoC\#2. This value comes very close to a 17 arcsec HEW derived from a direct surface metrology and a 16 arcsec HEW estimated from a measurement in UV light\cite{Civitani} . 

\begin{figure}[hbt]
	\centering
     \includegraphics[width = 0.52 \textwidth]{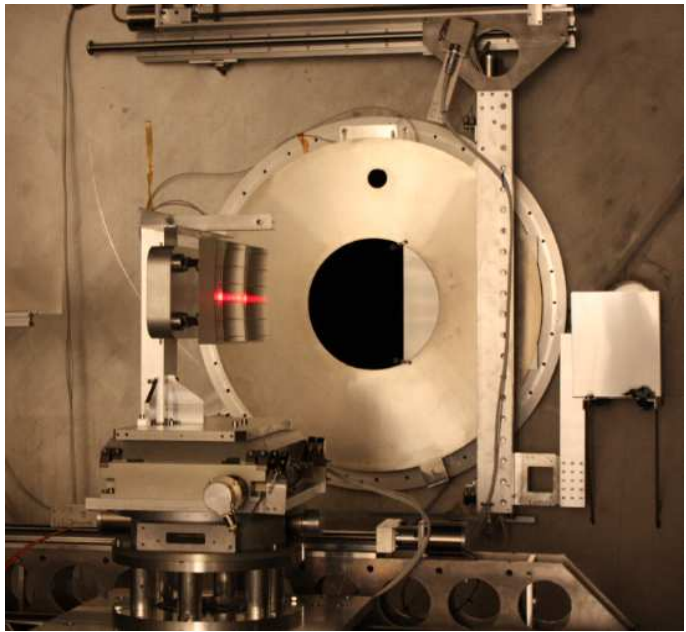}
	\caption{The PoC\#2 optic mounted in the vacuum tank at the PANTER facility. The X-rays impinge on the optic from the circular aperture in the background, partly obstructed to avoid the direct illumination of the detector. The optic under test is mounted on the PANTER manipulator and is initially aligned using a laser. }
	\label{fig:setup}
\end{figure}

\section{Intra-focus tests at PANTER}\label{tests}
The PoC\#2 (Fig.~\ref{fig:PoC2}) is a stack of 4 thin (0.4~mm) mirror (heretofore denoted as "plate") pairs made of 200 mm $\times$ 200~mm glass foils. They were formed by hot-slumping in cylindrical shape with a $R_0$ = 1~m curvature radius, then stacked onto a glass backplane, kept spaced by $\sim$2.9~mm thick stiffening glass ribs\cite{Civitani}. The Wolter-I\cite{VanSpey} shape is obtained by enforcing the parabola/hyperbola profile against precisely figured integration moulds before the ribs are glued during the integration process.\cite{CivitaniOE} The 4 layers are mounted with a common focus, located at a $f$ = 20~m distance from the parabola/hyperbola intersection plane (heretofore I.P.) and named PP0 (Plate Pair No. 0), PP1, PP2, and PP3 (a dummy one). Clearly, the PP0 is free not only from rib obstructions, but also from any off-axis obstruction from other plate pairs. 

The PoC\#2 mounted at PANTER\cite{PANTER1, PANTER2} is shown in Fig.~\ref{fig:setup}. The optic is fixed on the PANTER manipulator at the exit of the 123~m long, 1~m diam. vacuum tube. from which X-rays impinge on the optic, and aligned in rotation (accurate to 1~arcsec) and translation (accurate to 1~$\mu$m). The selection of the individual PP is obtained moving a mask with a thin slit (3~mm wide) in front of the vacuum tube. Also a thick slit (15~mm wide) was drilled in the mask to enable the simultaneous illumination of the entire PoC\#2. The reflected rays are finally imaged by one of the available detectors\cite{PANTERext}, mounted at a 7950 mm distance from the I.P. of the optic. 

\begin{figure}[hbt]
     \begin{tabular}{ll}
     \includegraphics[width = 0.48 \textwidth]{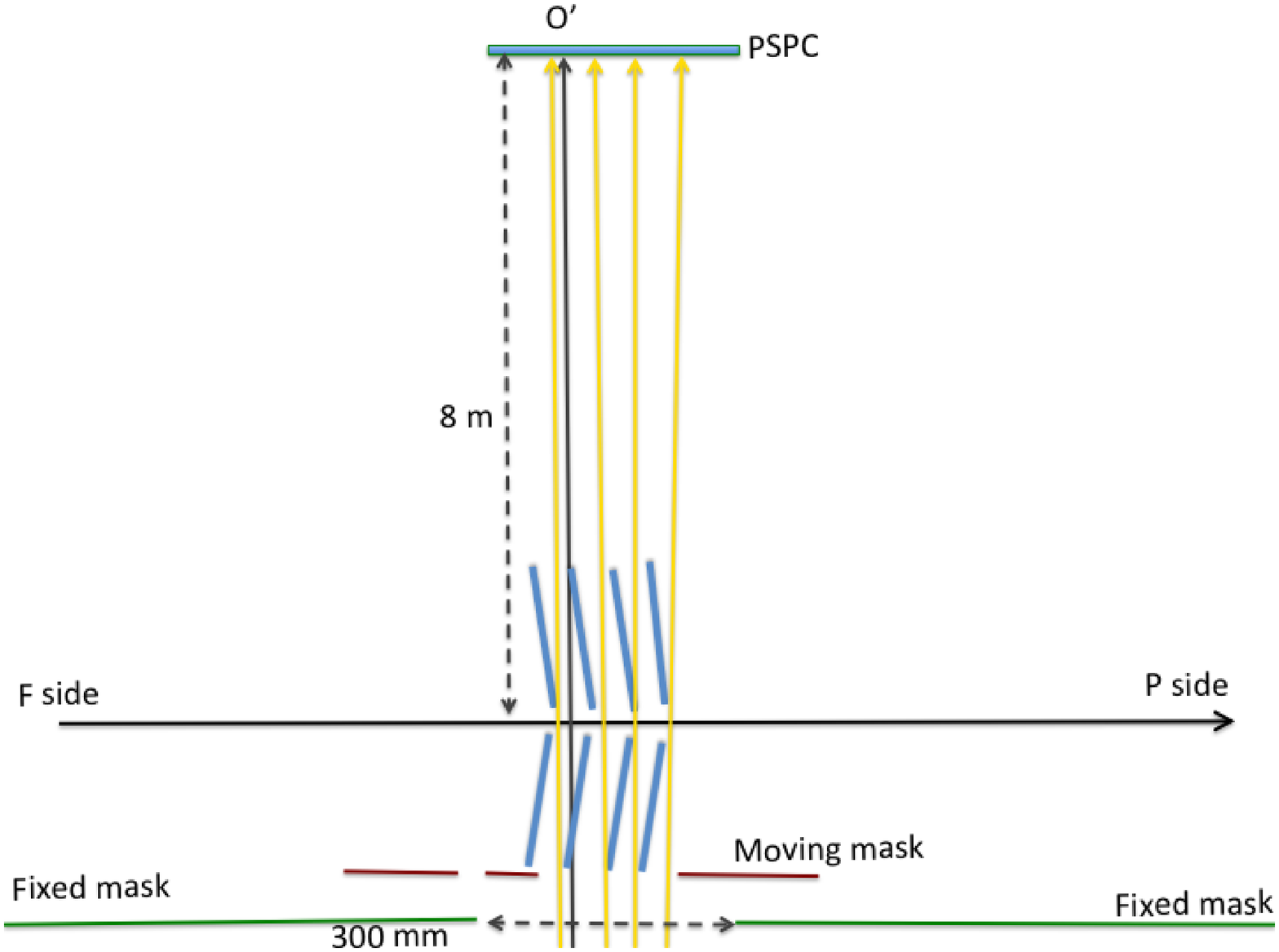}&
     \includegraphics[width = 0.48 \textwidth]{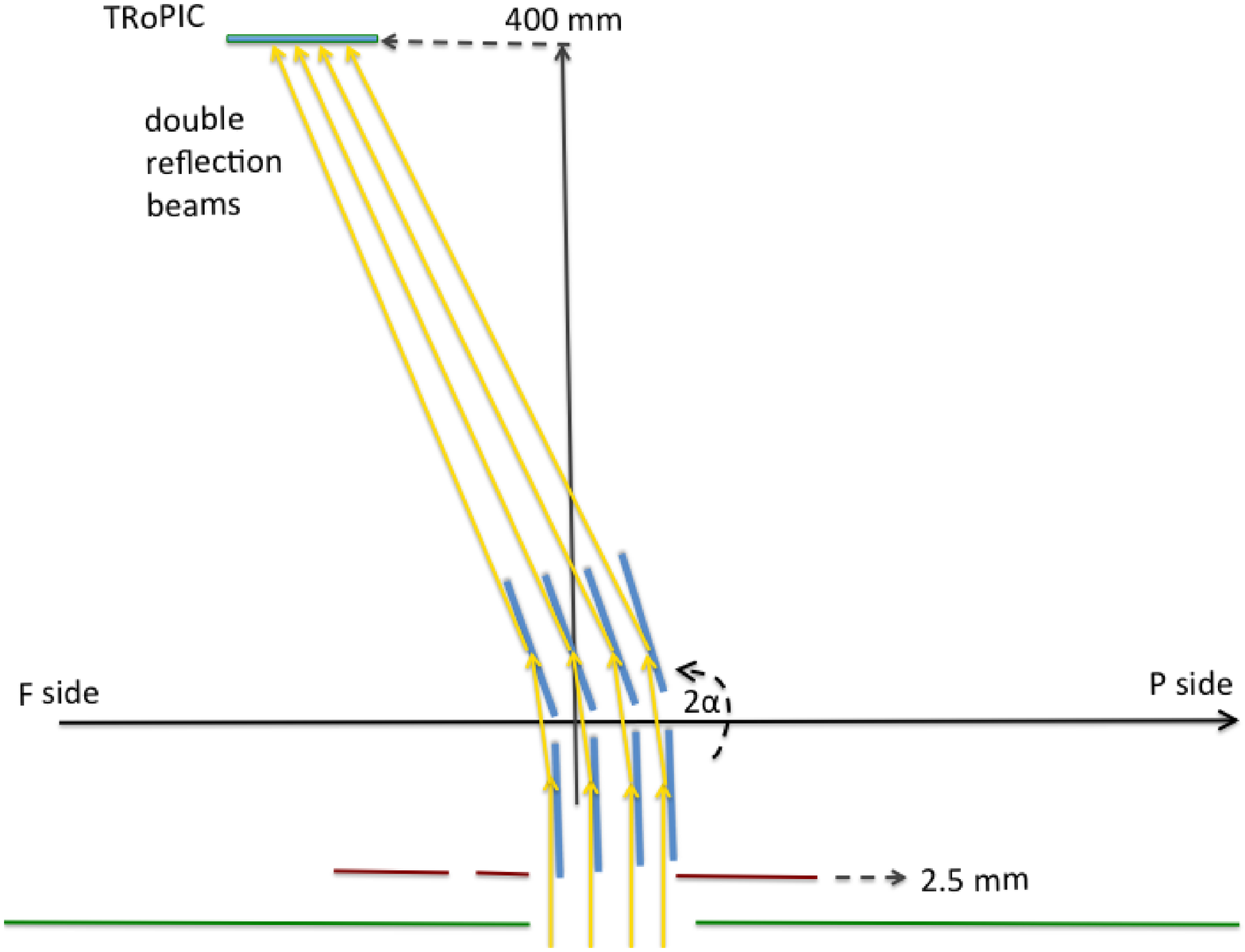}\\
     \footnotesize A) & \footnotesize B)
	\end{tabular}
	\caption{A) initial alignment of the PoC\#2 to the direct beam at PANTER, viewed from top. The zero of the angular scale is set maximizing the amplitude of the X-ray beam passing through the gaps between consecutive plate pairs. B) a 2$\alpha$-rotation of the PoC\#2 in counterclockwise sense sets the incidence angle at a nearly correct value for double reflection. The alignment is subsequently refined maximizing the intensity of the focused beam.}
	\label{fig:align1}
\end{figure}
\begin{figure}[hbt]
	\begin{tabular}{ll}
		\includegraphics[width = 0.48 \textwidth]{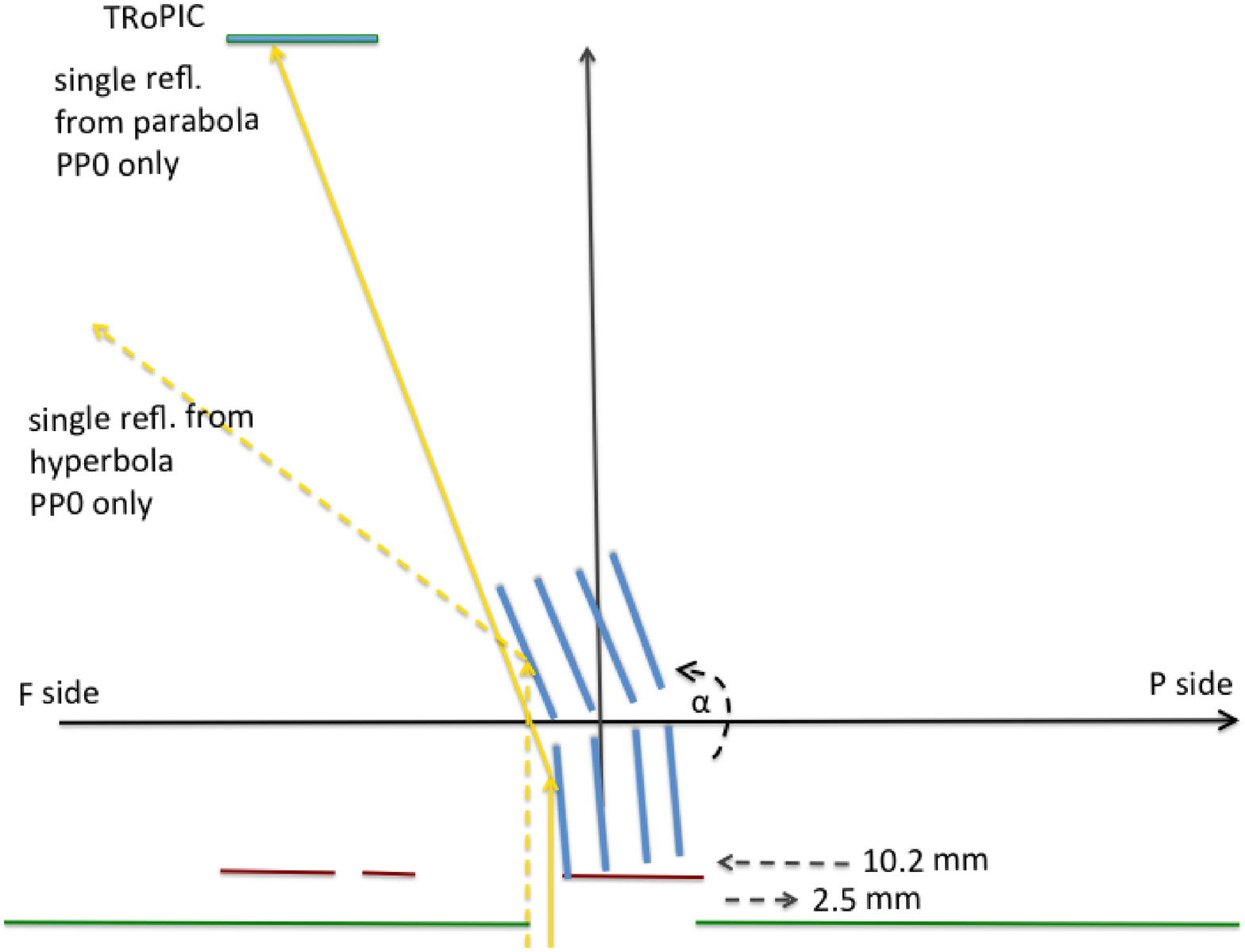}&
		\includegraphics[width = 0.48 \textwidth]{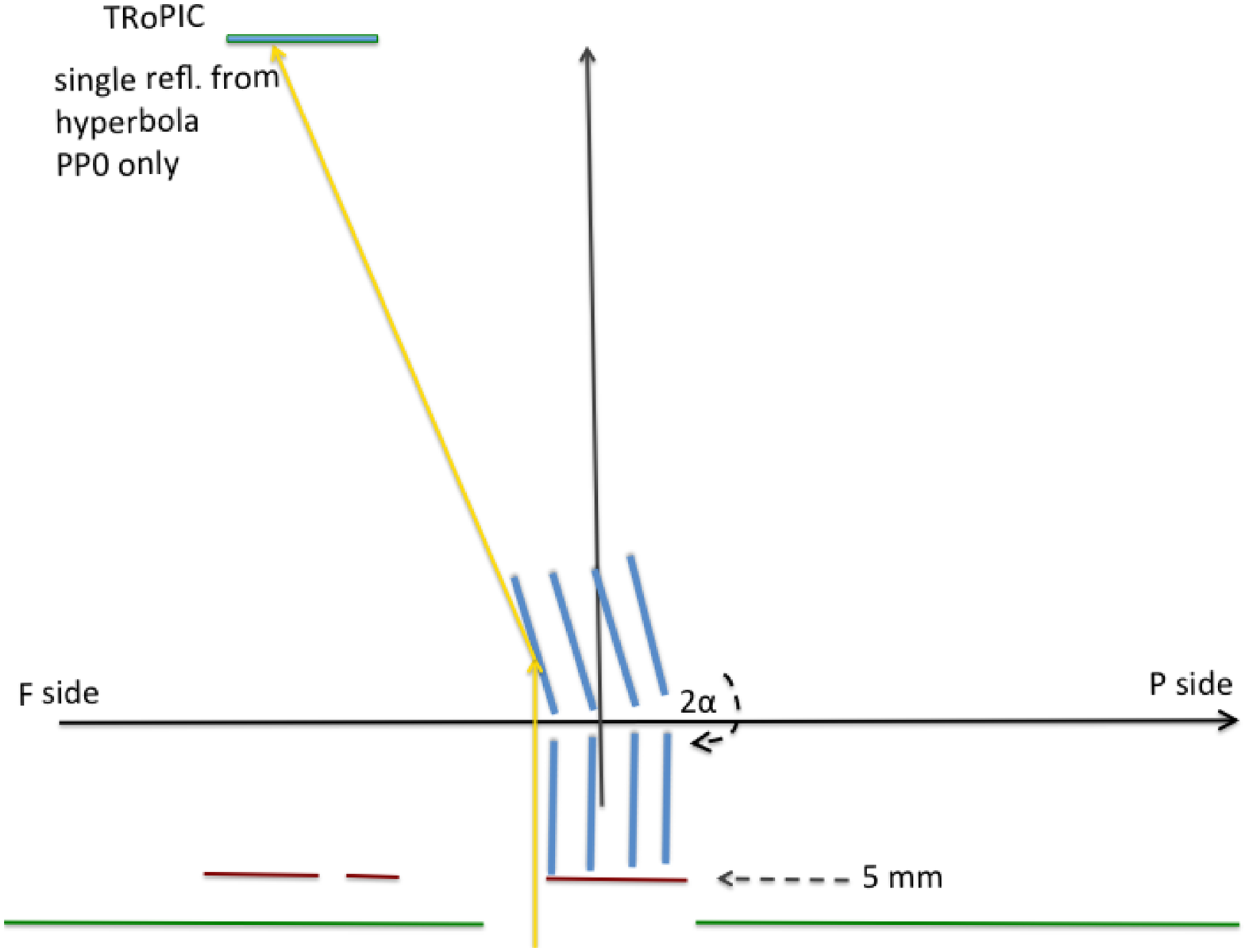}\\
	    \footnotesize A) & \footnotesize B)
	\end{tabular}
	\caption{A) The single reflection on the parabolic segment of the PP0 of the PoC\#2 is obtained via a further rotation by $\alpha$ in counterclockwise sense, viewed from top.  B) A subsequent rotation by 2$\alpha$ in clockwise sense sets the single reflection on the hyperbola of the PP0. In both cases, the reflection trace position remains almost unchanged, but the front mask position requires some adjustment. The inner plate pairs are obstructed by 80\% in single reflection setup.}
	\label{fig:align2}
\end{figure}

The PSPC (Position Sensitive Proportional Counter), a gas detector developed for the ROSAT telescope, has a circular sensitive area with a diameter of 8~cm, with moderate spectral (40\% at 0.93 keV) and spatial (250~$\mu$m) resolutions. Owing to its large field, it has been used for the initial alignment, even if the absence of visual references makes the focused beam search non-trivial. The adopted method consists of setting the PSPC in line with the POC\#2 and the X-ray source, then rotating it until X-rays pass through the gaps between consecutive plate pairs (Fig.~\ref{fig:align1}A). The rotation continues until the width of the passing beamlets reaches a maximum: this corresponds to a position of the optic in which the mirror edges at the entrance and exit pupils are aligned, and sets the zero of the rotation angular scale. The coarse alignment for double reflection is therefore obtained by rotating the PoC\#2 by 2$\alpha$, where $\alpha$~= 0.715~deg is the nominal incidence angle on-axis (see Fig.~\ref{fig:align1}B). In these conditions, even though the source is still divergent -- with the related effects\cite{PANTER2} like the focus displacement to $f'$ = 23.85~m -- the amount of rays that miss the second reflection is reduced to a negligible level and the setup mimics an on-axis source at infinity. Finally, the alignment was refined to within $\sim$0.5~arcmin maximizing the reflected count rate: that condition represents,  to a good approximation, the best alignment of the PoC\#2 in double reflection conditions.
 
Owing to their better imaging capabilities, the PIXI and the TRoPIC detectors\cite{PANTERext} have been used to scan the intra-focal images, that would ideally look like arcs 133~mm long and 1.7~mm thick (heretofore referred to as "traces"). In particular, TRoPIC was used to record in detail the intra-focal traces of the individual pair plates. TRoPIC is a solid state CCD camera with the same characteristics, but reduced size, of the imaging CCDs being produced for the eROSITA focal planes. TRoPIC is characterized by a smaller sensitive area (19.2~mm $\times$ 19.2~mm) than the PSPC, but higher spatial (75 $\mu$m pixel size) and spectral resolutions. The intra-focal traces are too long to entirely fit in the TRoPIC size. For this reason, they have been scanned with TRoPIC recording 14 images along the trace length. Every exposure lasted $\sim$15 min and overlapped the neighboring one by 50\% of its width: the images were subsequently assembled via software. The Carbon target was selected for the X-ray source; in addition to the C-K$\alpha$ line at 0.27~keV, also the 1.8~keV W-M$\alpha$ line from the Tungsten filament was present. Since the cutoff reflection energy at 0.715~deg for the bare glass is $\sim$2 keV, both X-ray lines are reflected. However, they can be easily separated by TRoPIC's excellent spectroscopic capabilities.
 
\begin{figure}[H]
\centering
\vspace{-10mm}
	\begin{tabular}{lllll}
     \includegraphics[height= 0.95 \textheight]{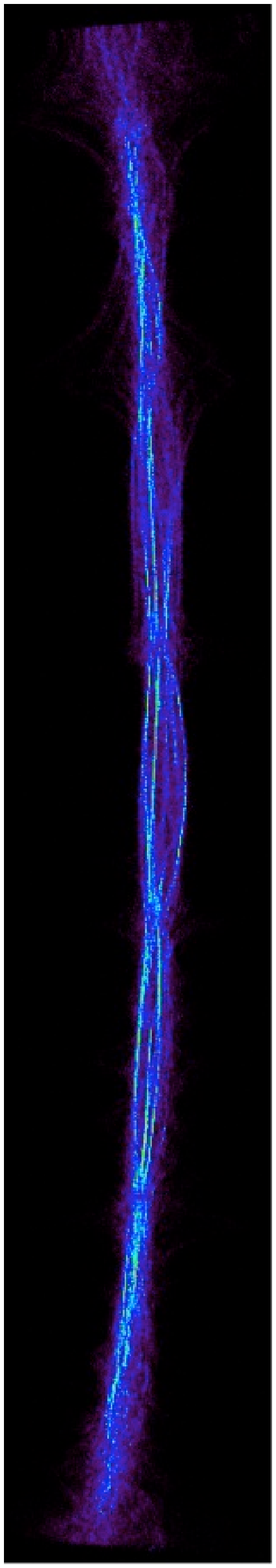}& 
     \includegraphics[height = 0.95 \textheight]{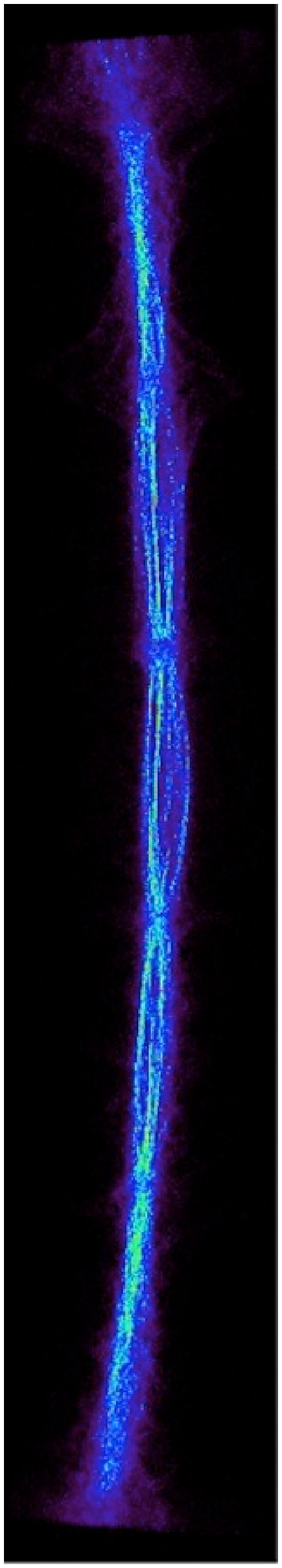}& 
     \includegraphics[height = 0.95 \textheight]{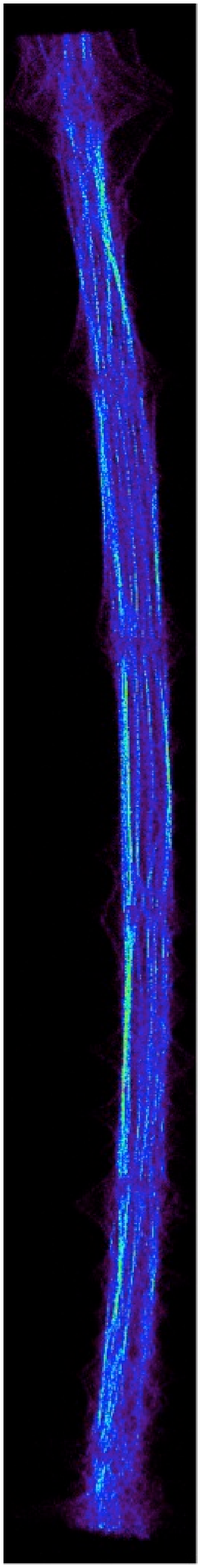}&
     \includegraphics[height = 0.95 \textheight]{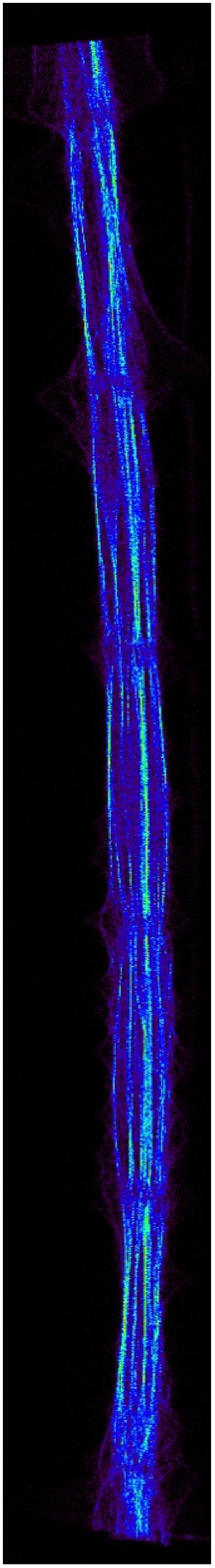} & 
      \hspace{-3mm}\includegraphics[height =  1.20\textwidth]{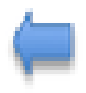}\\
     \footnotesize A) &\footnotesize B) &\footnotesize C) &\footnotesize D) &
     \end{tabular}
	\caption{Collected TRoPIC exposures on intra-focal traces of the PP0. Each panel has a 1.92~cm true width. A) double reflection at 0.27~keV. B) double reflection at 1.8~keV. C) single reflection on the parabola, 0.27~keV: the trace edge on {\it right} side corresponds to the mirror edge close to the I.P. D) single reflection on the hyperbola, 0.27~keV: the trace edge on {\it left} side corresponds to the mirror edge close to the I.P. The arrow on the right represent the position of the image slice taken as example in Fig.~\ref{fig:scheme}.}
	\label{fig:TRO_images}
\end{figure}

The intra-focal double reflection trace of the PP0 at 0.27~keV is shown in Fig.~\ref{fig:TRO_images}A, and at 1.8 keV in Fig.~\ref{fig:TRO_images}B. In the remainder of this work we concentrate the analysis on the PP0, while the other traces are shown in another work\cite{Civitani}. If the mirrors were perfect, the PP0 trace would be {\it a regular arc of 1.7~mm thickness and uniform brightness}. In contrast, mirror profile errors result in a trace with a more complicated pattern, which, under the conditions that will be listed in Sect.~\ref{recon}, provides {\it a detailed mapping of the mirror surface}. For example, the trace at 0.27~keV (Fig.~\ref{fig:TRO_images}A) exhibits {\it a width modulated by the periodicity of the rib underneath the glass}. The ribs locations are marked by a trace width close to 1.7~mm and a quasi-uniform intensity distribution: this means that the glass profile at the rib locations is very close to the nominal profile, i.e., {\it the shape of the accurately figured integration moulds has been replicated under the ribs}. The situation changes moving to the space between ribs, where the trace becomes {\it broader}, as a consequence of the glass spring-back that tends to restore the initial profile of the foil as-slumped, and {\it nonuniform} because of small oscillations in the longitudinal profile, having a typical period of a few centimeters. In fact, the brightness distribution is organized in nearly-parallel lines that appear to be "squeezed" at rib locations, where the mould shape is enforced. This scenario is in qualitative agreement with the expectations from Finite Element Analysis\cite{CivitaniOE}. Finally, the comparison of the images at 0.27~keV and 1.8~keV shows that the increase of the X-ray scattering is almost negligible, because the two traces are essentially the same. Both traces are sharp with a readable topography, excepting the two sides where the brightness pattern becomes confused. In other words, the roughness is within the tolerances -- as expected from roughness measurements of foils before the integration\cite{Civitani} -- everywhere but at the two ends in azimuthal direction.

In order to obtain quantitative information on the mirror profile, we need to measure the mirrors in single reflection off either the parabola or the hyperbola. This can be done easily as shown in Fig.~\ref{fig:align2}, respectively doubling or nulling the incidence angle on the parabola, and adjusting the entrance mask position. However, this was possible only for the PP0: the single reflections off the other PPs are heavily obstructed by the tight stacking of the mirrors. 

The single reflection traces, also 133~mm long, have been generated by the X-ray source with Carbon target and  scanned with TRoPIC in the same, aforementioned way: nevertheless, the 1.8~keV line is no longer reflected because the incidence angle was doubled. The single-reflection traces at 0.27~keV are shown in Fig.~\ref{fig:TRO_images}C (parabola) and Fig.~\ref{fig:TRO_images}D (hyperbola). Qualitatively, both of them exhibit the same features of the double reflection traces: their width is close to the nominal one (this time, 4.8~mm for the parabola and 4.5~mm for the hyperbola) at the sole rib location. The trace becomes gradually broader in the infra-rib space, where the glass spring-back prevails. The surface topography is affected by small mid-frequencies that modulate the brightness map in quasi-parallel lines, which are suppressed at locations corresponding to the ribs positions. Both traces are sharp but at the two ends, where the intensity lines are clearly scattered around by a locally higher roughness. But, additionally, in single reflection setup {\it the brightness distribution is closely related to the local curvature in the longitudinal direction}, hence it allows us to quantitavely reconstruct the part of the mirror profile that chiefly affects the optical quality. In the next section we expose the formalism that enables this computation.

\section{Longitudinal profile reconstruction}\label{recon}
\subsection{Basic assumptions and approximations}\label{approx}
\begin{figure}[H]
	\centering
	\begin{tabular}{ll}
	\includegraphics[width=0.59 \textwidth]{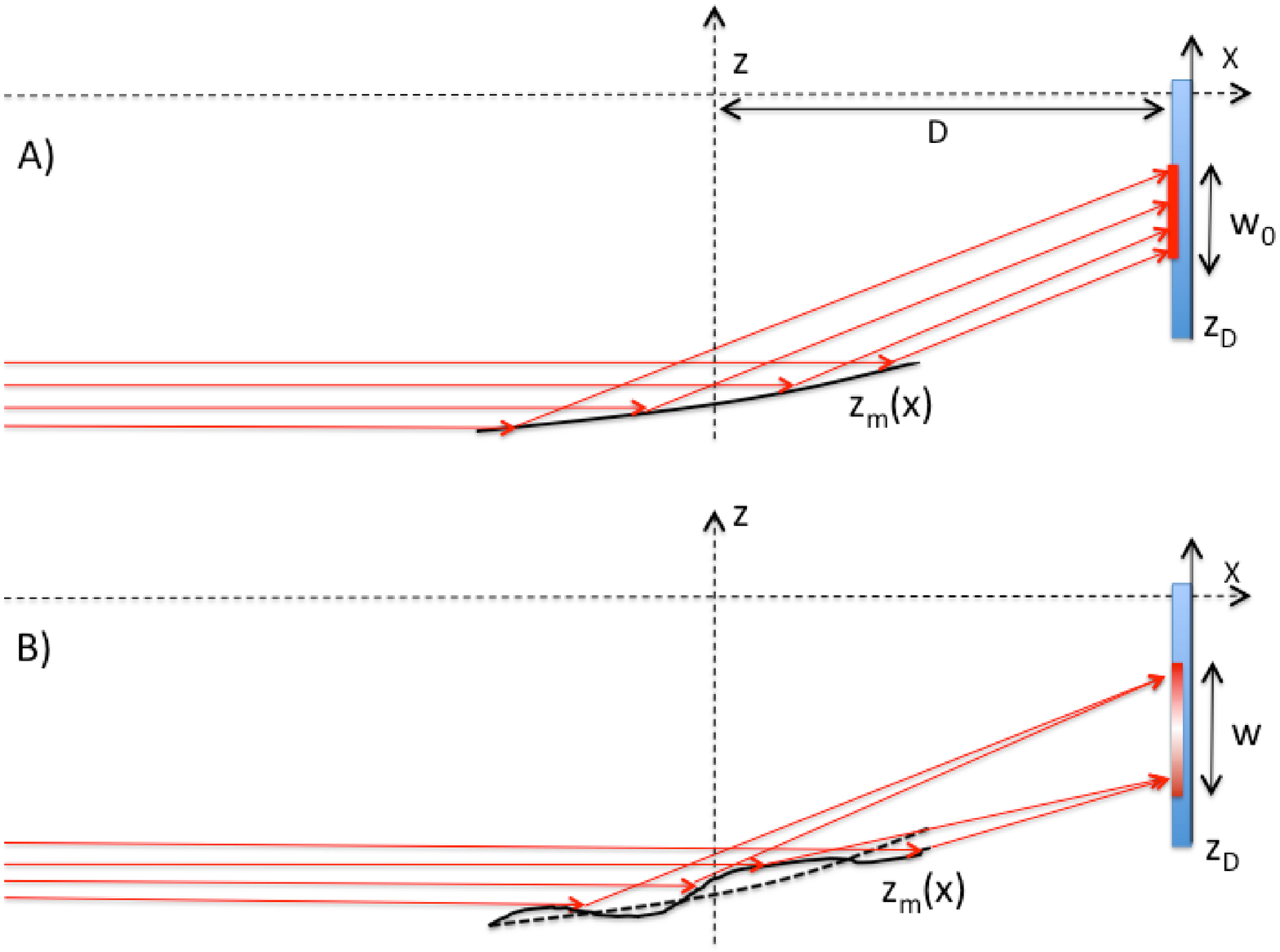} & 
	\includegraphics[width=0.38 \textwidth]{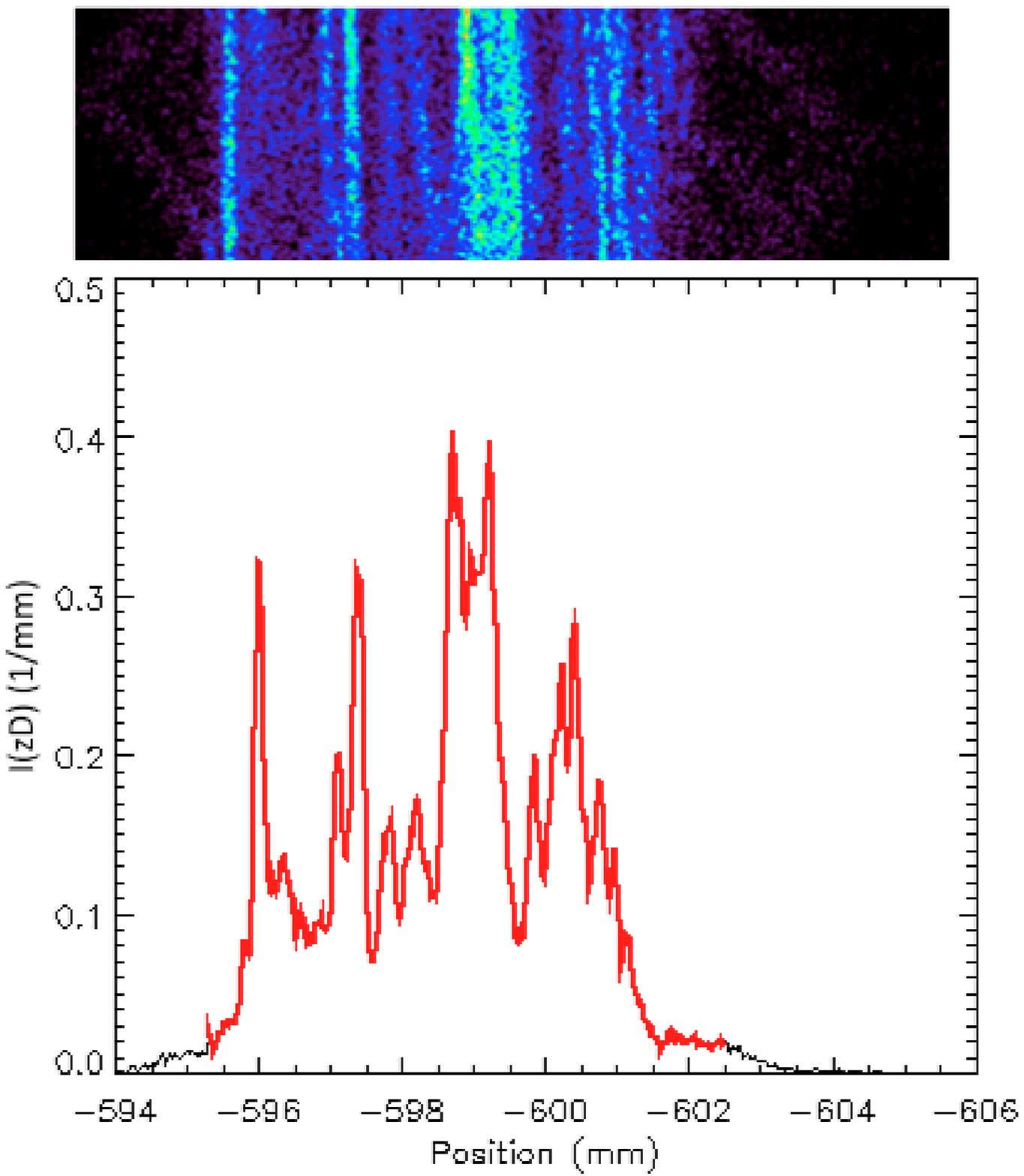}
	\end{tabular}
	\caption{Left side: the principle of the reconstruction of the profile error from the intensity observed on the detector in intra-focal position. The incidence angles are greatly exaggerated. A) a perfect mirror would return a uniform trace; B) the presence of defects affects both width and brightness variations of the trace. Right side: a slice of the intra-focal trace of the hyperbola single reflection (arrow in Fig.~\ref{fig:TRO_images}D). This slice corresponds to a 5 mm wide slice of the mirror, 35 mm off-center. The projected intensity along the $z_D$ axis is displayed in the viewgraph on right side.}
	\label{fig:scheme}
\end{figure}

The basic principle we adopt to reconstruct a mirror profile is shown in Fig.~\ref{fig:scheme}, left: {\it convex surfaces concentrate the light, concave surfaces spread it out}. Hence, the mirror profile curvature varies along the optical axis (the $x$ coordinate, toward the focus), yielding a local variation of the intensity along the direction locally orthogonal to the trace and oriented toward the PoC\#2 optical axis (the $z_D$ coordinate). We set the mirror center at $x$=0 and denote with $L$ the mirror length, with $R_0$ its azimuthal radius at the I.P., with $D$ the distance from the origin to the detector. We preliminarily assume that:
\begin{enumerate}
	\item{geometrical optics can be applied, i.e., interferential effects like X-ray scattering are negligible;}
	\item{the deflection out of plane caused by roundness errors is negligible;}
	\item{$D$ is {\it small enough to avoid the rays to cross each other}, but at the same time large enough to allow the reflected rays to converge or diverge;}
	\item{$D$ is much larger than the mirror length.}

\end{enumerate}

Condition 1) has to be fulfilled for the trace to map the mirror profile, i.e. to uniquely associate a point on the mirror to a point on the trace. In practice, at a given X-ray energy, it can be non-trivial to check {\it a priori} the absence of physical optics effect, i.e., which range of spatial wavelengths can be treated with geometrical optics if the surface is unknown\cite{RaiSpi1}. Therefore, the validity of the assumption has to be checked {\it after} the profile reconstruction, verifying that the predictions of the geometrical and the physical optics actually merge and return the measured trace (Fig.~\ref{fig:physical}A).

Condition 2) simply requires that the roundness errors are not so large to affect the distribution of the intensity along the $z_D$ coordinate. This is usually fulfilled in grazing incidence by good quality mirrors. In the present case, the condition is apparently met because the brightness lines are quasi-parallel and do not form loops. Roundness errors (especially far from the focal plane) are expected to affect the trace curvature in the azimuthal direction, therefore also roundness errors can be computed from the trace shape. However, we do not consider this aspect in this work.

Condition 3) is very important. It states that the detector-to mirror distance, $D$, {\it has to be optimized for the specific mirror quality} to enhance the intensity variations over the trace, but at the same time the crossing of rays reflected from different parts of the mirror {\it must} be avoided. Put in another way, the trace-mirror correspondence must be {\it unique}. In the present case, the condition is very likely fulfilled, because {\it the intensity lines do not cross each other}, which makes easy to collapse a thin slice of the measured trace into an brightness profile (Fig.~\ref{fig:scheme}, right). The situation changes, however, at both trace ends where the intensity distribution becomes confused because of roughness and -- probably -- also of the worse profile. Should the mirror have only a worse shape, it would be sufficient to reduce $D$ until no line crossing is seen, indeed at the expense of the intensity contrast (Fig.~\ref{fig:monotonic}).

Condition 4), finally, is needed to rule out obliquity effects, i.e., to allow us to make some approximations in the formalism explained in the next paragraph. 

\subsection{The relation between trace brightness and mirror profile}\label{curv_equation}
We consider a radial section of the intra-focal trace as in Fig.~\ref{fig:scheme}, right. Its distance --measured along the trace-- from the center is proportional to the distance of the corresponding mirror profile from the mirror center, so the azimuthal coordinate is determined. The intensity projection of the slice along the radial direction, per length unit, is denoted with $I(z_D)$. A ray propagating {\it along the x-axis} impinges on the mirror at the generic $x$ coordinate, is reflected at the generic incidence angle ${\hat\alpha}$, and reaches the detector at the coordinate $z_D$ at a distance $D$. If geometrical optics can be applied, $z_D$($x$) is provided by the following equation, {\it valid for shallow reflection angles}:
\begin{equation}
	z_D(x) = z_{\mathrm m}(x)+2z'_{\mathrm m}(x)(D-x),
	\label{eq:deviation}
\end{equation}
where $z_{\mathrm m}$ is the real mirror profile, including additional tilts, and the prime denotes a derivative. To simplify the notation, in the following we drop off the $z_{\mathrm m}$ explicit dependence on $x$. Eq.~\ref{eq:deviation} is the mapping relation between detector and mirror coordinate, and is an unknown quantity because $z_{\mathrm m}$ is unknown. Vice versa, were $z_D$ a known function, the straightforward solution of Eq.~\ref{eq:deviation} would return $z_{\mathrm m}$. For example, if we set $z_D(x) =0$ for all $x$ and $D=f$ in Eq.~\ref{eq:deviation}, we obtain by integration $z_{\mathrm m} = C \sqrt{f-x}$, where $C$ is a constant. Imposing the condition that $z_{\mathrm m}(0) = -R_0$ we obtain $C=-R_0 f^{-1/2}$. So we get
\begin{equation}
	 x = f- f \frac{z_{\mathrm m}^2}{R^2_0}.
	\label{eq:parab}
\end{equation}
Eq.~\ref{eq:parab} represents, as expected, a parabolic profile with axis along the negative $x$ direction. Indeed, its focus is located at $x =f(1-\tan^2{\hat\alpha})$, i.e., in practice slightly before the expected focal plane. This small discrepancy is caused by the approximation of the incidence angle with the profile derivative in Eq.~\ref{eq:deviation}: in practical cases, it amounts to a few millimeters, less than the uncertainty often introduced by figure errors.

 \begin{figure}[H]
	\begin{tabular}{lll}
     	\includegraphics[width=0.31 \textwidth]{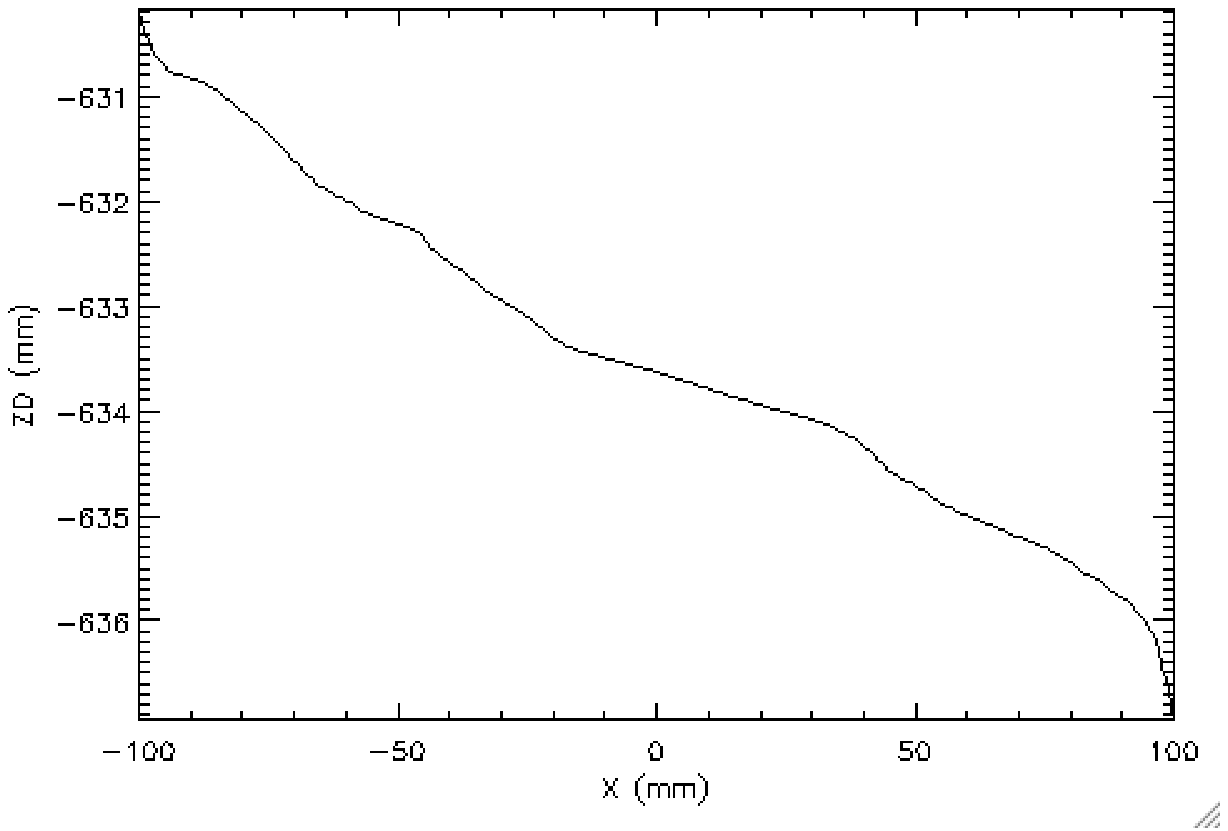} & 
		\includegraphics[width=0.31 \textwidth]{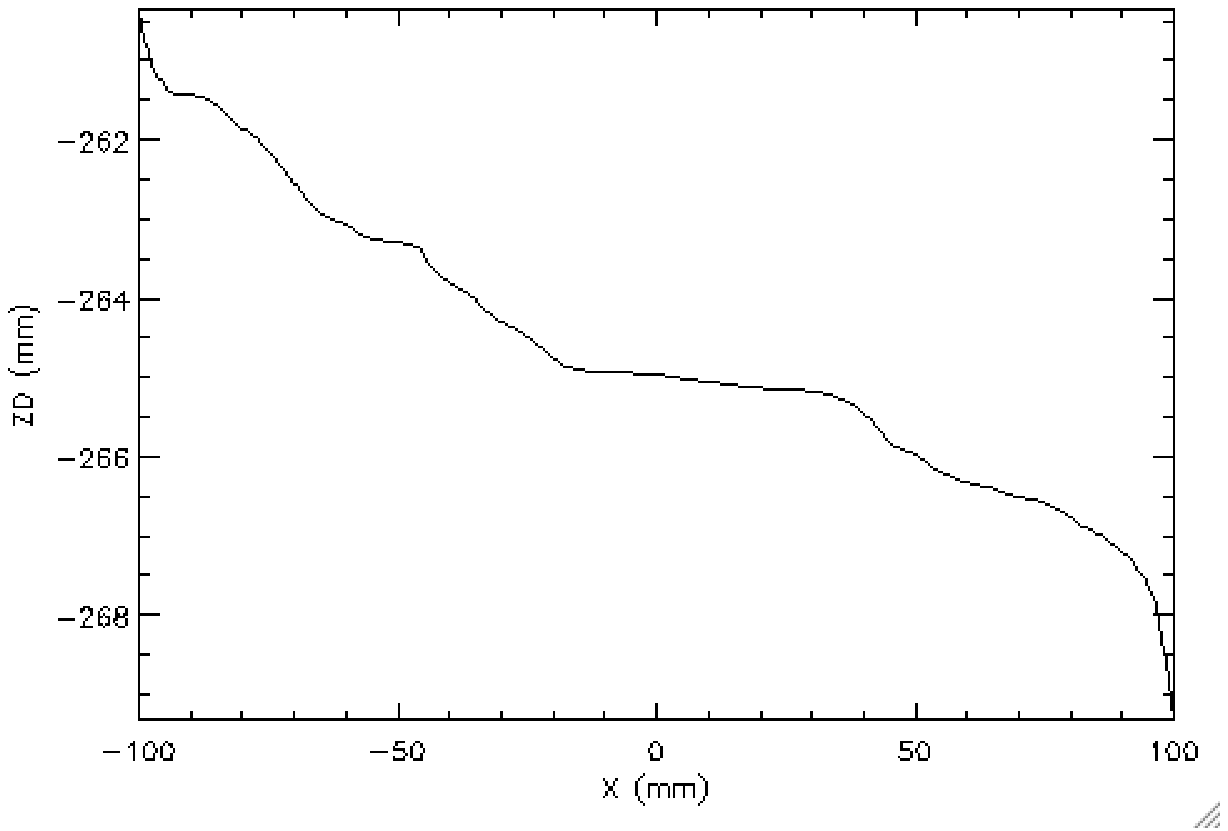} & 
		\includegraphics[width=0.31 \textwidth]{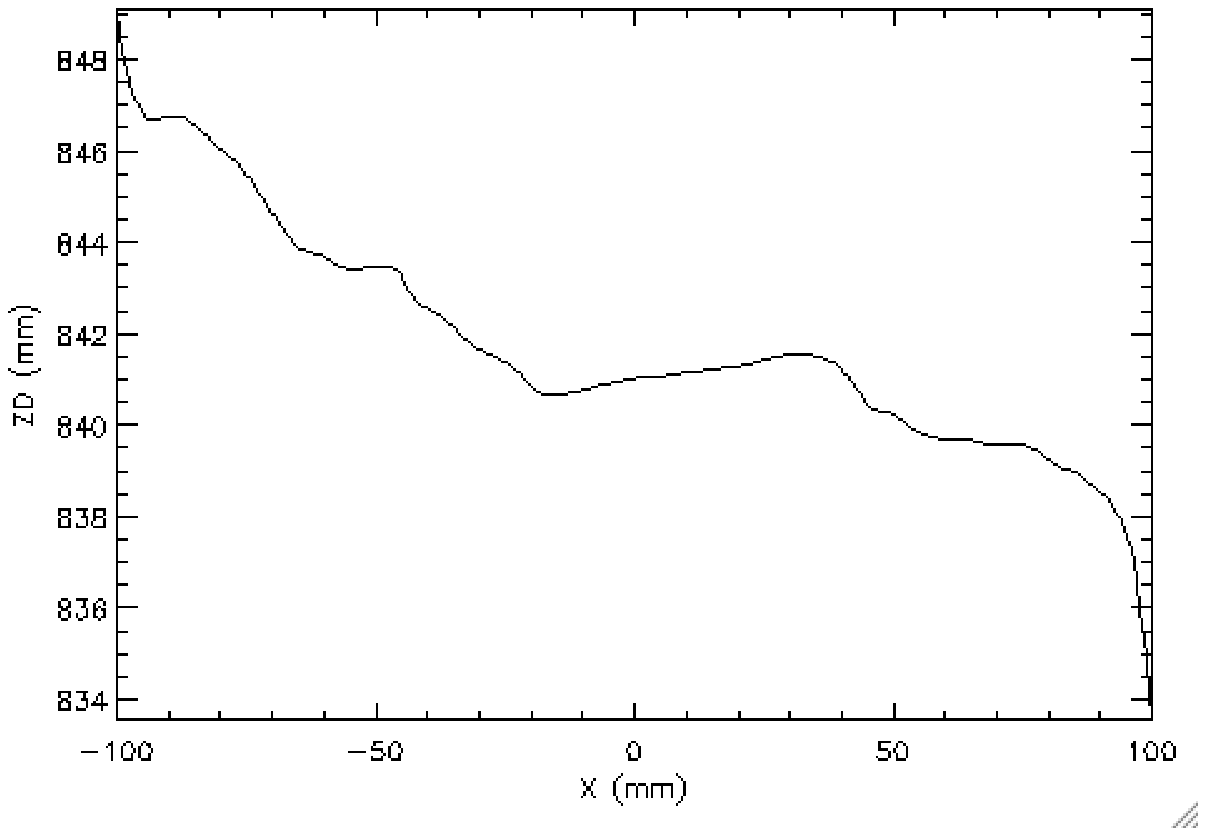}
     \end{tabular}
	\caption{Evolution of the $z_D(x)$ function with $D$, the mirror-detector distance, assuming as error profile the one shown in Fig.~\ref{fig:steps}. From left to right: D = 8~m, 16~m, and 40~m. The relation deviates from the linearity for increasing values of $D$, until it is no longer decreasing everywhere. We note that in single reflection measurement, because of the off-axis setup, the beam is not focused.}
	\label{fig:monotonic}
\end{figure}

For the correspondence $z_D \leftrightarrow x$ to be unique, as requested by condition 3), Eq.~\ref{eq:deviation} must be monotonic. As already mentioned in Sect.~\ref{approx}, this occurs if $D$ is not too large. The example shown in Fig.~\ref{fig:monotonic} shows that the closer to the mirror, the more the slope errors get amplified and the relation $z_D(x)$ becomes complicated. For $D$~= 8~m, the function is still decreasing everywhere and the following equations can be applied. We now differentiate Eq.~\ref{eq:deviation}:
\begin{equation}
	\Delta z_D = 2(D-x)\,\Delta z'_{\mathrm m}-z'_{\mathrm m}\Delta x.
	\label{eq:deviation_diff}
\end{equation}
Because the function in Eq.~\ref{eq:deviation} is decreasing, $\Delta z_D < 0$.  As the incident beam is initially uniform, the collected intensity by a mirror element of length $\Delta x$ is proportional to $z'_{\mathrm m}\Delta x$, but it is concentrated/spread over a $|\Delta z_D|$ length on the detector. Hence, the image brightness at $z_D$ is
\begin{equation}
	I(z_D) = I_0\frac{z'_{\mathrm m}\Delta x}{|\Delta z_D|} = \frac{I_0 z'_{\mathrm m}}{z'_{\mathrm m}-2(D-x)\, z''_{\mathrm m}},
	\label{eq:intensity}
\end{equation}
where $I_0$ is a constant, representing the radiation intensity impinging on the mirror. We thereby obtain a differential equation for the mirror slope, 
\begin{equation}
	 \left(1- \frac{I_0}{I(z_D)}\right) \, z'_{\mathrm m} = 2(D-x)\, z''_{\mathrm m}.
	\label{eq:intensity_diff}
\end{equation}
A similar approach to solve a beam-shaping problem was already presented by one of the authors\cite{Spiga13}. To understand the meaning of Eq.~\ref{eq:intensity_diff}, we derive a simplified form approximating $z'_{\mathrm m} = z'_{\mathrm m}(0)$ and $D-x \approx D$,
\begin{equation}
	 z''_{\mathrm m}(x) \simeq \frac{z'_{\mathrm m}(0)}{2D}\left(1- \frac{I_0}{I(z_D)}\right):
	\label{eq:deri2}
\end{equation}
from Eq.~\ref{eq:deri2} it appears that the locations corresponding to a beam concentration ($I(z_D) >  I_0$) are concave upwards ($ z''_{\mathrm m} >0$), while the ones corresponding to a beam dispersion ($I(z_D) <  I_0$) are concave downwards ($z''_{\mathrm m}<0$). If the focused beam intensity is uniform, the mirror curvature is also. This confirms the initial expectations.

Defining the {\it modulation function} $M(z_D) = 1- I_0/I(z_D)$, the solution of Eq.~\ref{eq:intensity_diff} is
\begin{equation}
	z'_{\mathrm m}(x) = z'_{\mathrm m}(0) \exp\left[\int_{0}^{x}\frac{M(z_D)}{2(D-t)}\,\mbox{d}t\right],
	\label{eq:solution}
\end{equation}
where $z'_{\mathrm m}(0)$, the mirror slope at $x =0$, is a {\it multiplicative} constant to be determined by constraining the section to comply the nominal one,
\begin{equation}
	\int_{-L/2}^{+L/2} z'_{\mathrm m}(x) \, \mbox{d}x = L\sin{\hat\alpha}.
	\label{eq:1st_cond}
\end{equation}
Therefore, if the integral in Eq.~\ref{eq:solution} can be solved, the mirror profile is obtained via a further integration of $z'_{\mathrm m}$ over $x$. In Eq.~\ref{eq:1st_cond}, ${\hat\alpha}$ is the real incidence angle on the mirror. In the case of the PoC\#2 optic, a Wolter-I one, ${\hat\alpha} = 2\alpha$ if the mirror is correctly aligned for single reflection (Fig.~\ref{fig:align2}). However, if the mirror is misaligned the constant $z'_{\mathrm m}(0)$ will be affected by an error that -- since the entire slope function is multiplied by its value -- will add a spurious curvature to the computed profile, rather than an unessential tilt. A particular attention has therefore to be paid to the mirror alignment.

Finally, we have to determine the normalization constant $I_0$. To this end, we define $w$, the total width of the trace (see Fig.~\ref{fig:scheme}B), and we compute it applying Eq.~\ref{eq:deviation} at the edges of the profile:
\begin{eqnarray}
	 w & = & \left[z_{\mathrm m}\!\left(\!-\frac{L}{2}\right)+2z'_{\mathrm m}\!\left(\!-\frac{L}{2}\right)\left(D+\frac{L}{2}\right)\right]-\left[z_{\mathrm m}\!\left(\frac{L}{2}\right)+2z'_{\mathrm m}\!\left(\frac{L}{2}\right)\left(D-\frac{L}{2}\right)\right] \approx \nonumber\\
	 &\approx& -Lz'_{\mathrm m}(0)+L\left[z'_{\mathrm m}\!\left(\frac{L}{2}\right)+z'_{\mathrm m}\!\left(-\frac{L}{2}\right)\right]- 2D\left[z'_{\mathrm m}\!\left(\frac{L}{2}\right)-z'_{\mathrm m}\!\left(-\frac{L}{2}\right)\right]\approx\nonumber\\
	 &\approx& Lz'_{\mathrm m}(0)- 2D\left[z'_{\mathrm m}\!\left(\frac{L}{2}\right)-z'_{\mathrm m}\!\left(-\frac{L}{2}\right)\right].
	\label{eq:width_form}
\end{eqnarray}
We now integrate Eq.~\ref{eq:deri2} over the mirror length and obtain
\begin{equation}
	 z'_{\mathrm m}\!\left(\frac{L}{2}\right)-z'_{\mathrm m}\!\left(-\frac{L}{2}\right) = \frac{z'_{\mathrm m}(0)}{2D}\left[L-I_0\int_{-L/2}^{+L/2}\frac{1}{I(z_D)}\,\mbox{d}x\right]\approx \frac{z'_{\mathrm m}(0)L}{2D}\left[1-I_0\left\langle \frac{1}{I(z_D)}\right\rangle\right],
	\label{eq:deltasl}
\end{equation}
where in the last passage of Eq.~	\ref{eq:deltasl} we have anticipated the relation $x \simeq - (L/w)z_D$ introduced in the next section (Eq.~\ref{eq:location}). Substituting this result into Eq.~\ref{eq:width_form} and solving for the constant $I_0$ we have
\begin{equation}
	 I_0 = \frac{w}{Lz'_{\mathrm m}(0)}\left\langle \frac{1}{I(z_D)}\right\rangle^{-1},
	\label{eq:I0}
\end{equation}
that provides the $I_0$ constant in terms of measurable quantities.

\subsection{Algorithm for profile reconstruction}\label{algo}
\begin{figure}[H]
\centering
	\begin{tabular}{ll}
        	 \includegraphics[width = 0.95\textwidth]{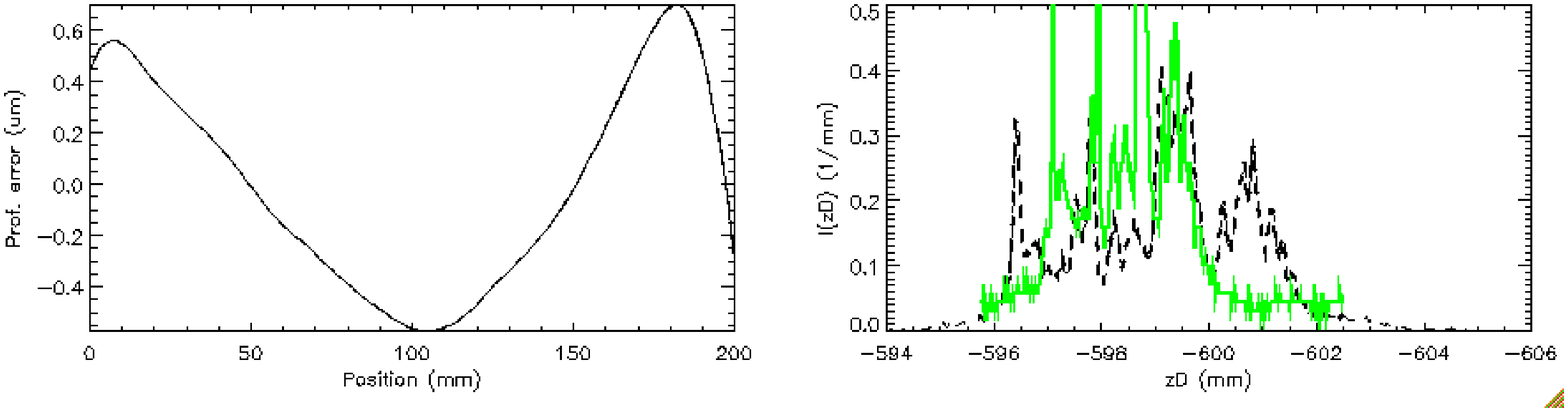} &\hspace{-3mm}\footnotesize A)\\
    	     \includegraphics[width = 0.95 \textwidth]{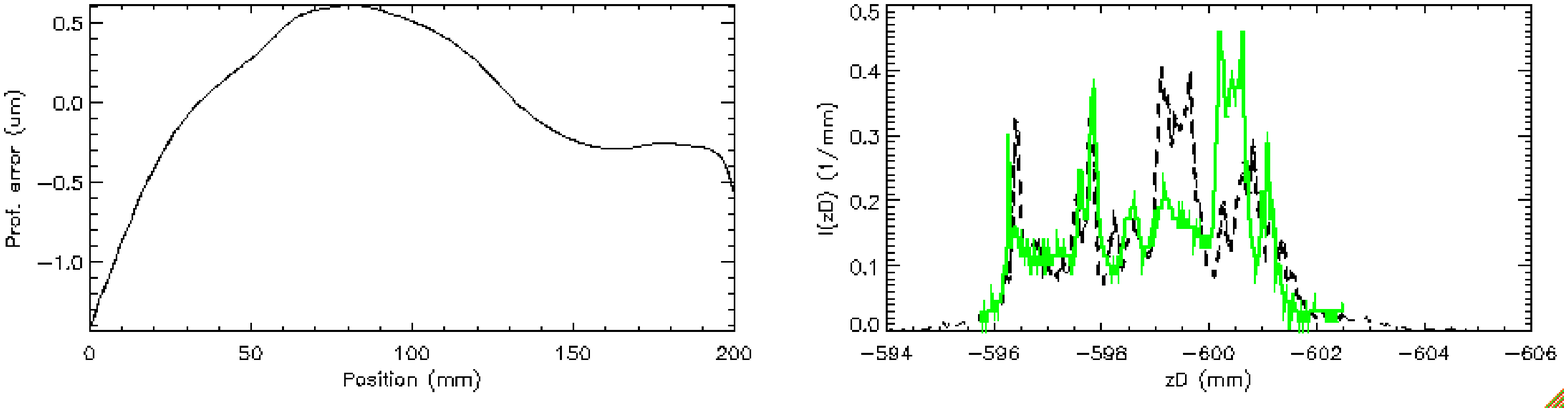} &\hspace{-3mm}\footnotesize B)\\
    	     \includegraphics[width = 0.95 \textwidth]{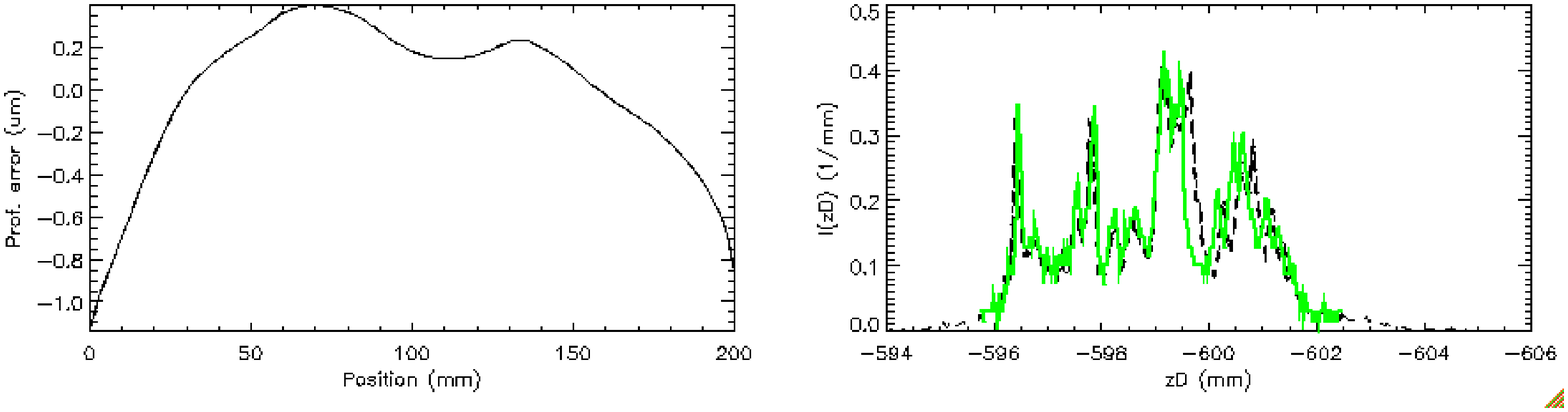} &\hspace{-3mm}\footnotesize C)\\
	     \includegraphics[width = 0.95 \textwidth]{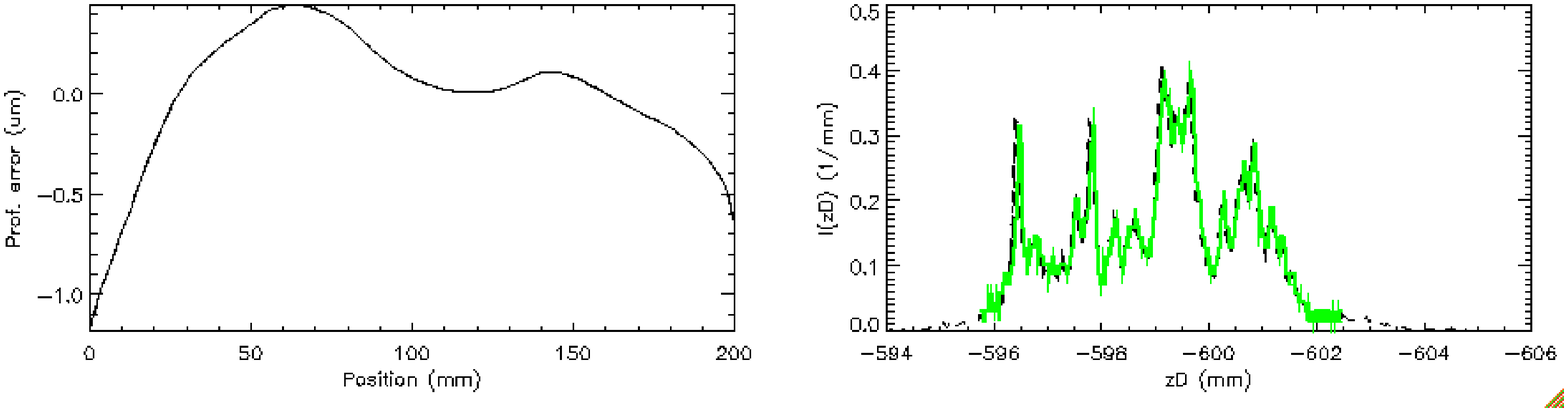} &\hspace{-3mm}\footnotesize D)
     \end{tabular}
	\caption{Left column: successive approximations of the mirror profile obtained from the intensity distribution in Fig.~\ref{fig:scheme} by iterative application of Eq.~\ref{eq:solution}. Right column: the intensity distribution computed from the respective mirror profile (solid line) compared with the measured one (dashed line). A) 1$^{\mathrm{st}}$ iteration. B) 2$^{\mathrm{nd}}$ iteration. C) 4$^{\mathrm{th}}$ iteration. D) 20$^{\mathrm{th}}$ iteration. Further iterations do not change the profile noticeably.}
	\label{fig:steps}
\end{figure}

Eq.~\ref{eq:solution} provides the derivative of the mirror profile $z_{\mathrm m}(x)$ corresponding to the brightness profile measured by the imaging detector, $I(z_D)$. However, the integral that appears in it cannot be directly solved. The reason is that $I$ depends on the coordinate $z_D$, while the integration has to be carried out with respect to $x$. Hence, the integration requires one to have in hand the explicit relation $z_D(x)$, which in turn requires (Eq.~\ref{eq:deviation}) to know $z_{\mathrm m}(x)$. To exit the loop, we approach the problem by subsequent approximations: an example of this procedure, applied to the brightness profile in Fig.~\ref{fig:scheme}, is displayed in Fig.~\ref{fig:steps}.

\begin{enumerate}
\item{We initially assume the relation between $x$ and $z_D$ to be a simple proportionality:
\begin{equation}
	x \simeq -\frac{L}{w}z_D,
	\label{eq:location}
\end{equation}
where for simplicity we have set the origin of $z_D$ at the center of the intensity profile. Moreover, assume $z'_{\mathrm m}(0) = \sin{\hat\alpha}$.}
\item{Use Eq.~\ref{eq:location} to solve the integral in Eq.~\ref{eq:solution}, then derive $z'_{\mathrm m}(x)$. A further integration provides $z_{\mathrm m, 0}$, a zero-order solution for an unmodulated intensity profile (Fig.~\ref{fig:steps}A, left).}
\item{Compute the corresponding $z_{D, 0}(x)$ via Eq.~\ref{eq:deviation}, then construct a histogram of the obtained values and compare it with the measured intensity profile (Fig.~\ref{fig:steps}A, right).}
\item{If the comparison returns a poor accord, then {\it resample} the measured $I(z_D)$ over the computed $z_{D, 0}(x)$.}
\item{Use the resampled $I(z_{D,0})$ in Eq.~\ref{eq:solution} and derive a more refined profile of the mirror, $z_{\mathrm m, 1}$ (Fig.~\ref{fig:steps}B, left).}
\item{Compute the corresponding $z_{D, 1}(x)$ via Eq.~\ref{eq:deviation}, then construct a histogram of the obtained values and compare it with the measured intensity profile (Fig.~\ref{fig:steps}B, right).}
\item{The comparison should now be more satisfactory. Resample the measured $I(z_D)$ over the computed $z_{D, 1}(x)$.}
\item{Repeat steps 5, 6, and 7 to obtain successive refinements of the mirror profile: $z_{\mathrm m, 2},  z_{\mathrm m, 3}, \ldots,  z_{\mathrm m, N}, \ldots$ until the computation converges (Fig.~\ref{fig:steps}D).}
\end{enumerate}

In order to ease the convergence at every (say, $N^{th}$) step, before resampling the intensity at the new coordinates $z_D$, it is convenient to enforce the computed $z_{D, N}(x)$ to exactly cover the same range as the {\it measured} coordinates, adding to the $z_{\mathrm m, N}(x)$ profile a small polynomial correction, 
\begin{equation}
	z_{\mathrm corr, N} = \frac{\Delta z_{\mathrm D, N}-w}{4DL}\,x^2
	\label{eq:poly}
\end{equation}
where $\Delta z_{D, N} = \max(z_{D, N}) - \min(z_{D,N})$, and then re-compute $z_{D, N}(x)$. If the algorithm is properly implemented, the amplitude of the corrective polynomial   approaches zero after a few iterations. 

\section{Longitudinal profile reconstruction from intra-focus image}\label{profile}
\begin{figure}[hbt]
	\begin{tabular}{cc}
		\hspace{-4mm}
         	 \includegraphics[width = 0.50 \textwidth]{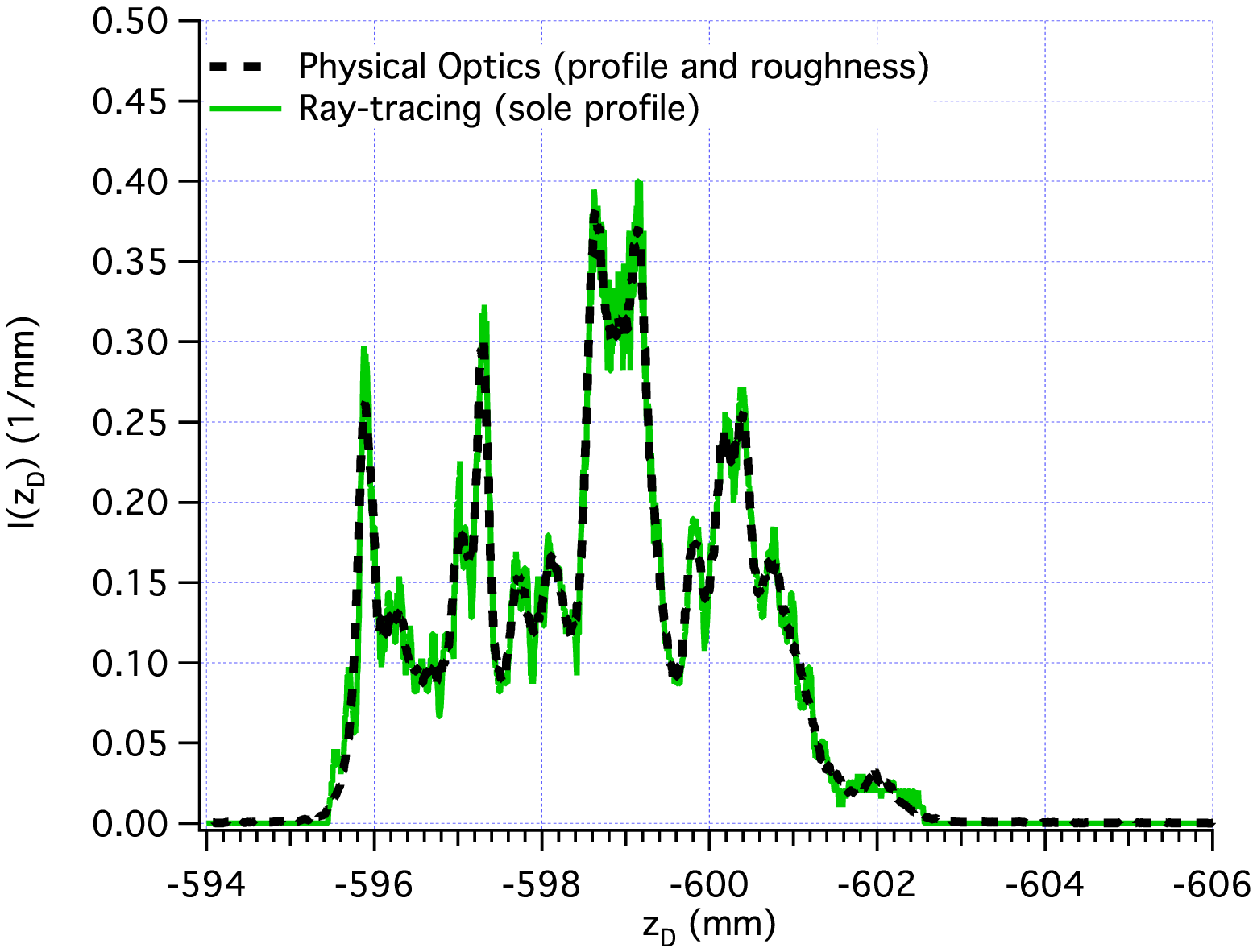}& 
     	 \includegraphics[width = 0.50 \textwidth]{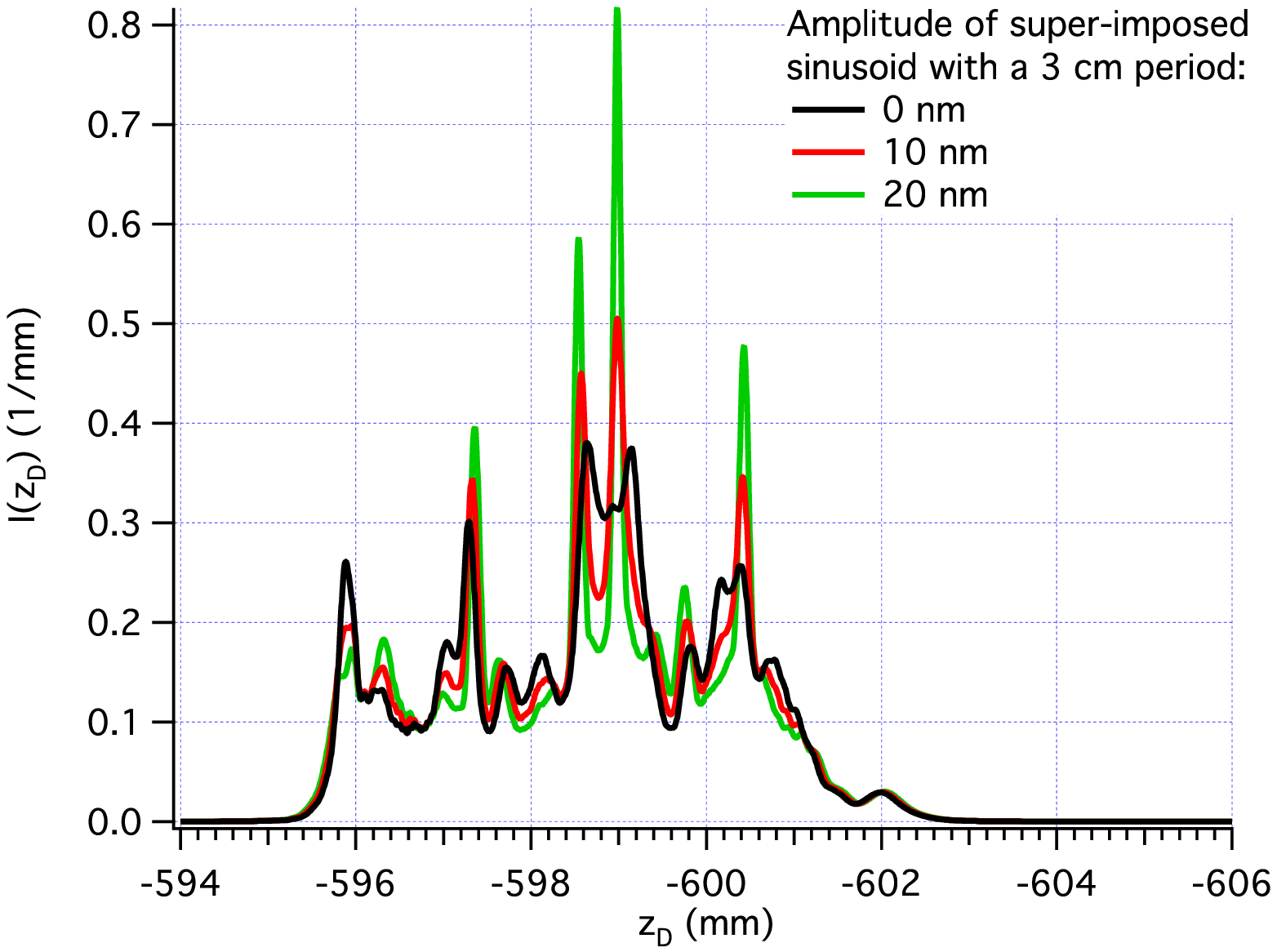}\\
	      \vspace{-1mm}
	      \footnotesize A) &\footnotesize B)
     \end{tabular}
     \vspace{1mm}
	\caption{A) Comparison of the profile intensity computed from geometrical optics and physical optics. The accord is excellent. B) Superimposing a waviness with a 3 cm spatial period to the measured profile would return a different intensity profile, depending on the perturbation amplitude. The spatial resolution of the computed profile is the same of the TRoPIC detector.}
	\label{fig:physical}
\end{figure}

We have applied the method developed in Sect.~\ref{recon} to the measured, single-reflection intra-focal traces of the PP0 shown in Fig.~\ref{fig:TRO_images}. The images were sliced into radial sections normal to the traces, neglecting thereby the azimuthal errors, up to a 70~mm distance from the median line. Beyond this limit, the intensity profiles become confused and cannot be used. The slice width corresponds to a 5 mm width on the mirror. The resolution of the longitudinal reconstructed profile is the one of the TRoPIC detector (75~$\mu$m) projected on the mirror longitudinal profile. However, it should be kept in mind that a sampling on the detector by $\Delta z_D$~= 75~$\mu$m steps corresponds, even at the zero-order approximation, to a step on the mirror larger by a factor of $L/w$ (Eq.~\ref{eq:location}). Since $w$ varies from profile to profile, the mirror sampling also varies. In the case we are considering, $w$ spans in the range between $\sim$4.8~mm at ribs and $\sim$7~mm amid consecutive ribs, yielding a sampling  2~mm $<\Delta x< $ 3.3~mm. This means that the reconstructed profile will include all spatial wavelengths larger than 4~mm (the Nyquist's), a range in which geometrical optics {\it could} be applicable at 0.27~keV, depending on the amplitude of the defects\cite{Aschenbach}. The validity of this assumption has, anyway, to be checked after the profile reconstruction by computing the expected $I(z_D)$ function using physical optics\cite{RaiSpi1, RaiSpi2}, and comparing it to the measured one (Fig.~\ref{fig:physical}A).
\begin{figure}[H]
\centering
	\begin{tabular}{cc}
	    \includegraphics[width = 0.44 \textwidth]{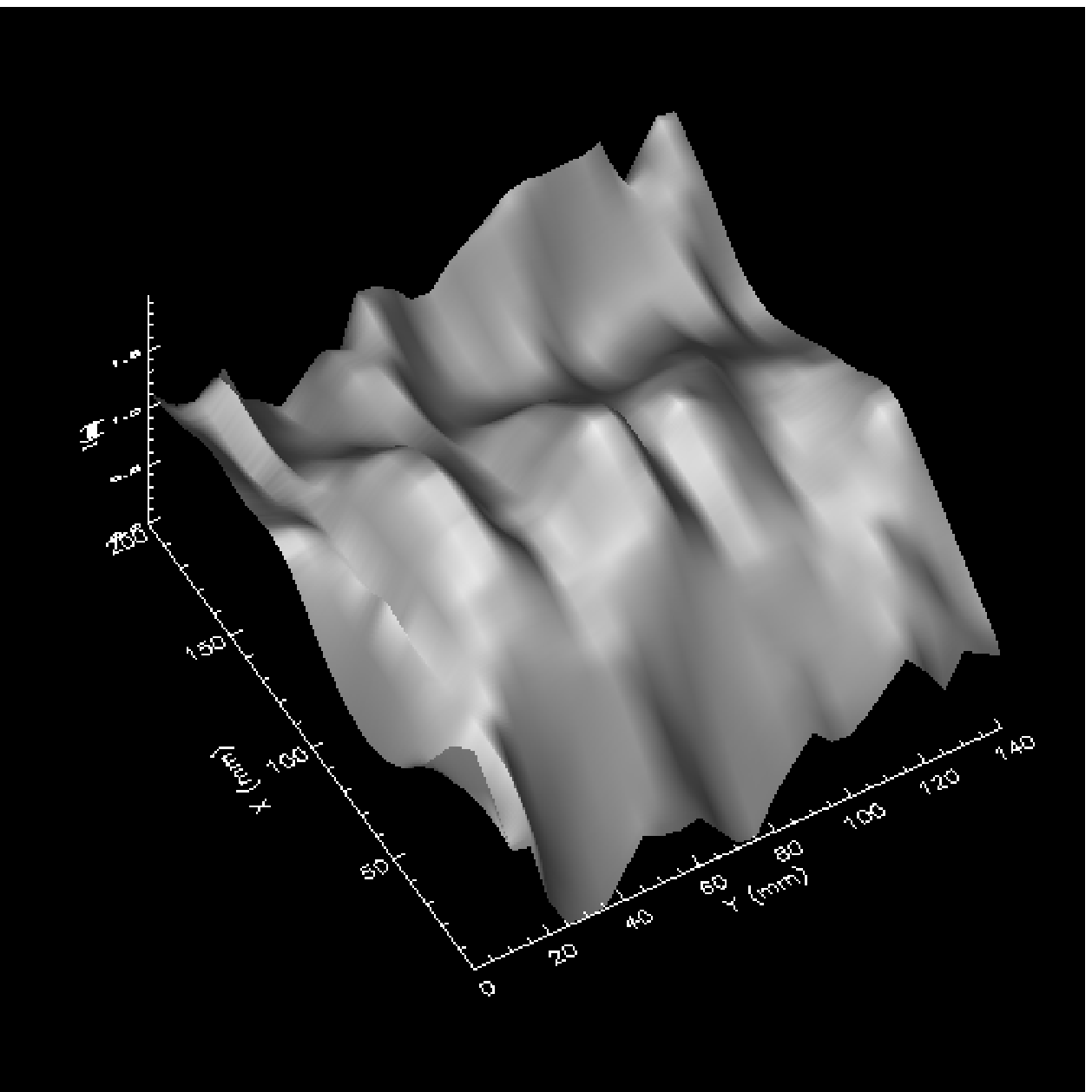}& 
        \includegraphics[width = 0.44 \textwidth]{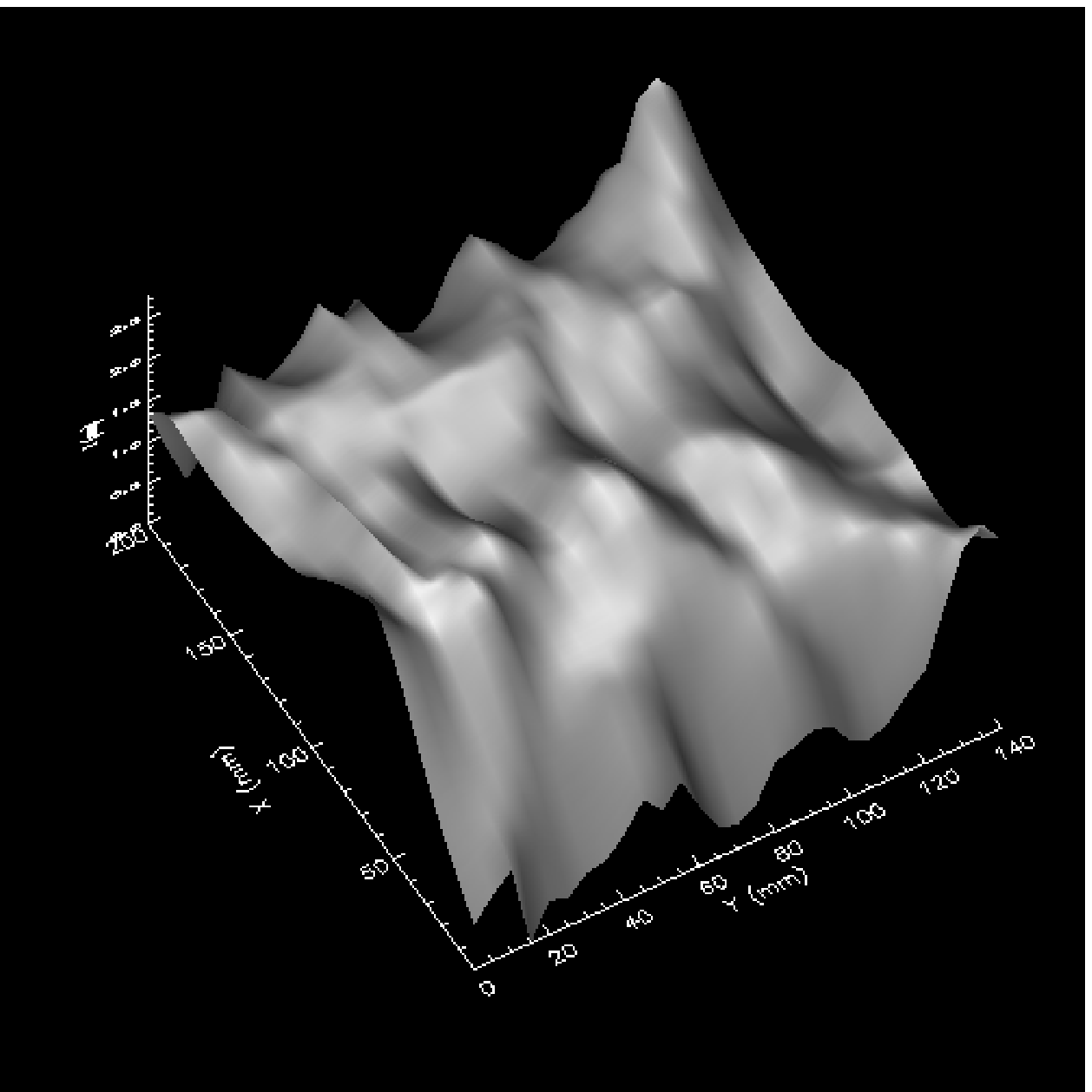}\\
        \footnotesize A) &\footnotesize B)
     \end{tabular}
	\caption{3D profile error (shaded map) of the A) parabolic segment and B) the hyperbolic segment of the PP0, as reconstructed from the intra-focal images. The roundness errors were not computed. In the parabolic map, the $x$-axis is oriented from the median to the maximum diameter. In the hyperbolic map, it is oriented from the median diameter toward the minimum diameter. As in Fig.~\ref{fig:scheme}, the $z$-axis is oriented toward the PoC\#2 optical axis.}
	\label{fig:3D}
\end{figure}

After slicing the trace, the brightness map has been collapsed onto the $z_D$ axis (Fig.~\ref{fig:scheme}B). Since the measured $I(z_D)$ is not perfectly sharp at the edges, we have set an intensity threshold to a few percent of its maximum. The remaining "tails" correspond to the very edges of the longitudinal profiles and {\it might} be associated to a higher roughness because of edge effects during the slumping process. 

\begin{figure}[H]
	\centering
	\begin{tabular}{lll}
     	 \includegraphics[width = 0.24 \textwidth]{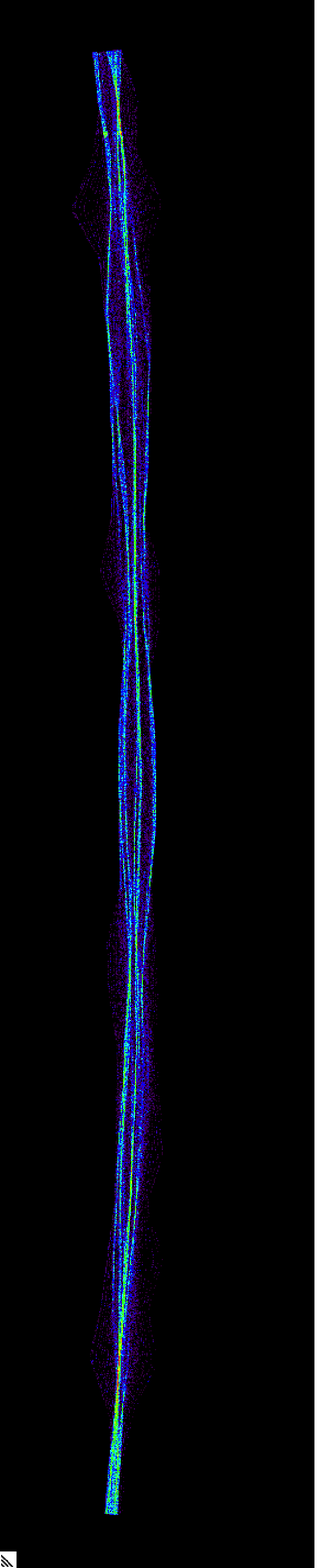}& 
     	 \includegraphics[width = 0.24 \textwidth]{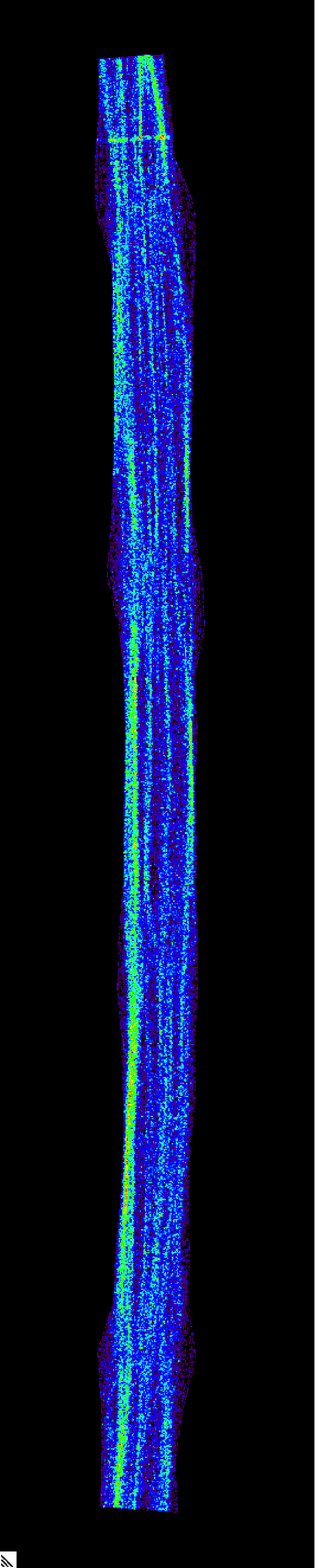}& 
     	 \includegraphics[width = 0.24 \textwidth]{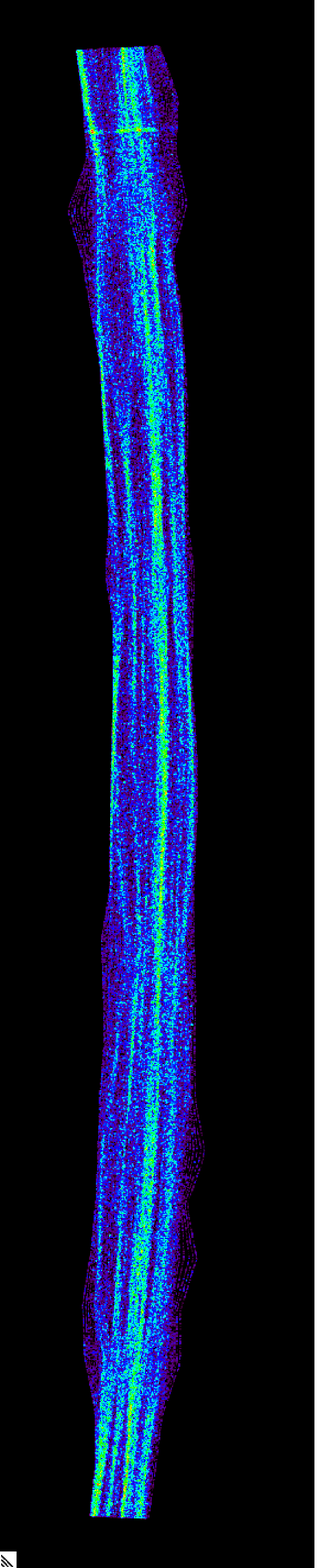}\\
          \footnotesize A) &\footnotesize B) &\footnotesize C)
	\end{tabular}
	\caption{Ray-tracing computation from the reconstructed part of the mirror surfaces (Fig.~\ref{fig:3D}). A) double reflection, B) single reflection on the parabola, C) single reflection on the hyperbola. Compare these intensity patterns with the measured ones in Fig.~\ref{fig:TRO_images}.}
	\label{fig:rt3D}
\end{figure}

Finally, the intensity profile has been processed as per the procedure described in Sect.~\ref{algo}, from the computed mirror profiles we have subtracted the nominal shape (either parabolic or hyperbolic), and the residuals have been assembled into two residual maps (Fig.~\ref{fig:3D}). Doing this, we have assumed no transverse offset or tilt related to possible azimuthal errors, because we disregard -- in this paper -- possible azimuthal errors (that were, indeed, later detected by direct mapping with the CUP profilometer at INAF/OAB\cite{Civitani}, but with a minor effect on the HEW in focus).

\begin{figure}[hbt]
	\centering
	\begin{tabular}{cc}
	\hspace{-4mm}
     	 \includegraphics[width = 0.50 \textwidth]{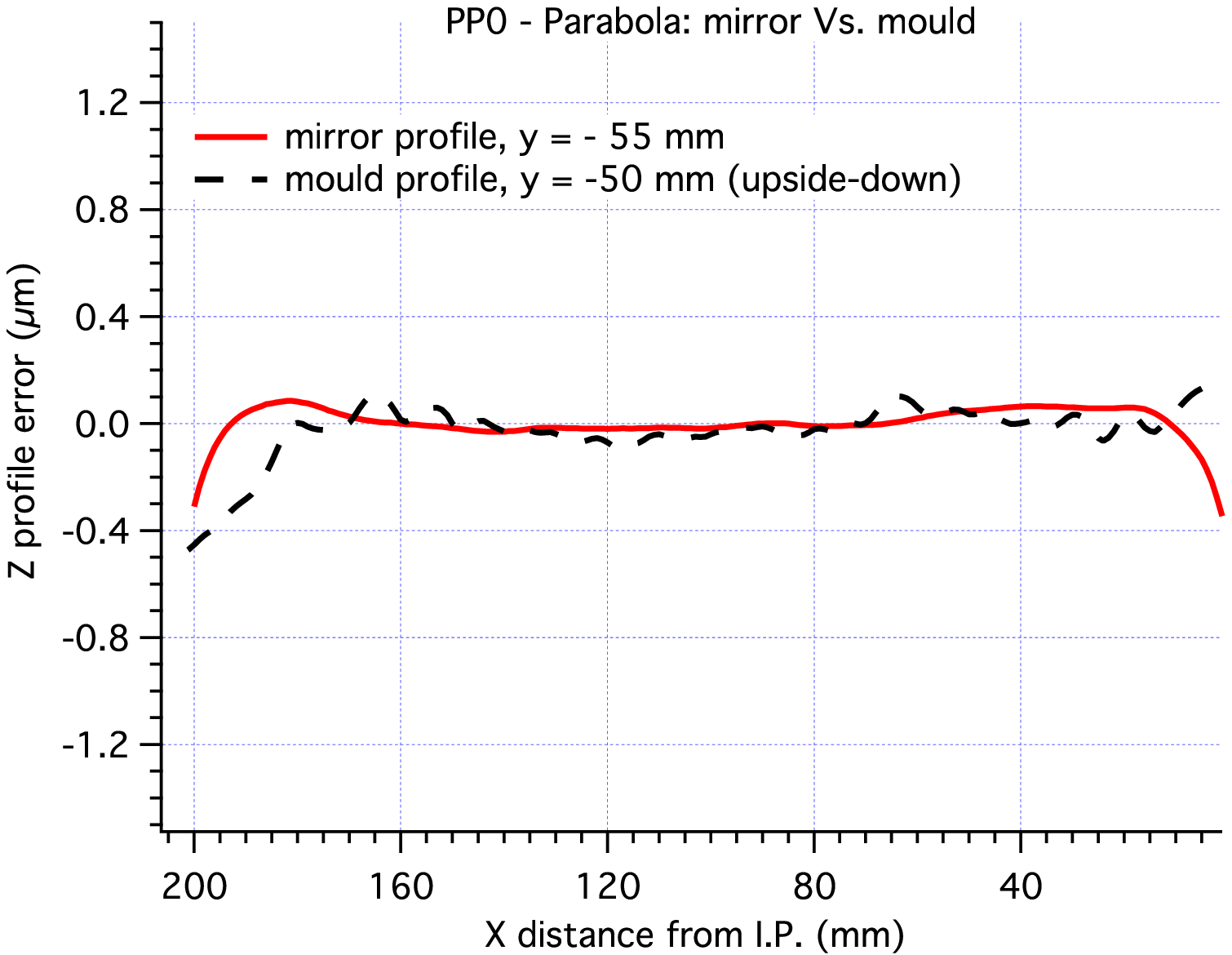}& 
     	 \includegraphics[width = 0.50 \textwidth]{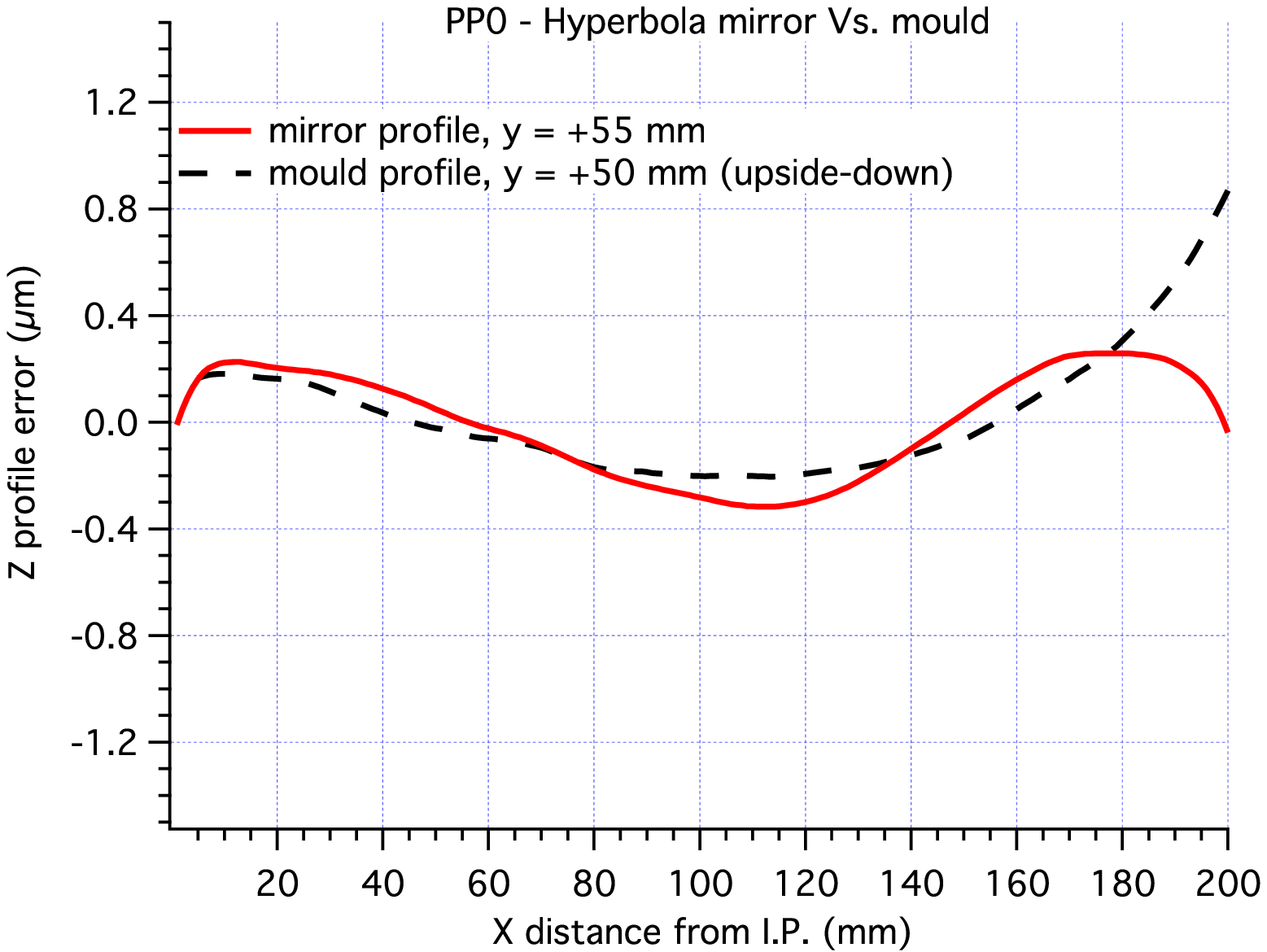}\\
	      \vspace{-1mm}
	      \footnotesize A) &\footnotesize B)
     \end{tabular}
     \vspace{1mm}
	\caption{Comparison between the profile error reconstructed from the intensity map (solid lines) of the PP0 near a rib location and an integration mould profile, as measured with the LTP, at a nearby $y$ coordinate (dashed lines). A) a parabolic mirror profile. B) a hyperbolic mirror profile.}
	\label{fig:ribs}
\end{figure}

The longitudinal profiles qualitatively exhibit the expected properties: the transverse modulation superimposed by the ribs is clearly visible. As already mentioned, the ribs behind the glass plate locally constrain the mirror to replicate the shape of the integration mould: in fact, the profile comparison in Fig.~\ref{fig:ribs} proves that the profile of the mould was effectively copied by the glass mirror. Only at the edges, the glass spring-back still tends to bend downwards because the ribs are slightly shorter than the mirror length, and the selected $y$ coordinate is not exactly in correspondence of a rib. Between the ribs, in contrast, the spring-back causes the glass to retain a small, but detectable part of its initial shape. Finally, the profiles exhibits very low oscillations over a 4-5~cm lateral scale and amplitude below 0.1~$\mu$m. These are responsible for the intensity modulations in Fig.~\ref{fig:TRO_images}.

The first check of the reconstructed maps in Fig.~\ref{fig:3D} is done by means of a ray-tracing routine in double and single reflection. The simulation results are displayed in Fig.~\ref{fig:rt3D}. The comparison with the double reflection and single reflection intra-focal traces, limited to the part comprised within the 4 central ribs (Fig.~\ref{fig:TRO_images}), shows an excellent agreement. The comparison would probably even better if the lateral sampling would have been tighter than 5~mm on the mirror. The only discrepancy is in the curvature of the traces, which could not be reproduced because the azimuthal errors were ignored in the profile reconstruction.

Another verification of the initial assumptions can be simply done re-computing the expected intra-focal intensity $I(z_D)$ using the physical optics, of general validity\cite{RaiSpi1}, instead of a ray-tracing routine based on geometric optics, from the reconstructed profile. The physical optics approach also allows us to account the scattering effect of the surface roughness. The result is shown in Fig.~\ref{fig:physical}A:  the exact computation from the mirror profile error, reported in Fig.~\ref{fig:steps} and including the measured surface roughness\cite{Civitani} (dashed line), is perfectly superposed to the ray-tracing findings from the sole reconstructed profile (solid line). This confirms the applicability of the condition 1) in Sect.~\ref{approx}: the geometrical optics can be applied to the profile information included in the intra-focal trace. Moreover, the surface roughness has a negligible effect -- at least, at this X-ray energy and this incidence angle -- on the brightness distribution of the intra-focal trace. This also confirms the expectations, because the measured roughness\cite{Civitani} fits the reference Power Spectral Density (PSD) established using a method described in another paper\cite{Spiga07}.

The last check regards the real absence of mid-frequencies in a spatial range below 4~cm. In fact, they might be present but the profile reconstruction might be -- {\it in principle} -- unable to detect them if very low, or they might not obey geometrical optics. To remove this doubt, we have repeated the physical optics simulation after superposing to the profile error under test an {\it additional} sinusoidal perturbation with a 3~cm period and decreasing amplitude. The resulting intensity distribution functions are displayed in Fig.~\ref{fig:physical}B, for 3 possible amplitudes of the perturbation: 0~nm, 10~nm, and 20~nm. The black line refers to the unperturbed profile and is the same shown in the panel A) of the same figure. The lines in color, simulated assuming a perturbation of 10 or 20~nm amplitude, yield a sensitively different intensity distribution along the $z_D$ coordinate, with more pronounced peaks that would be surely observed, if present, with the TRoPIC detector. We conclude that mid-frequencies errors at a 3~cm wavelength {\it must} have an amplitude of less than 10 nm. Similar simulations can be carried out to set an upper limit to higher frequency perturbations.  

\section{Conclusions and expected angular resolution}\label{concl}
In this paper we have described a method to directly derive the longitudinal profile of a mirror from a full-illumination, intra-focal, X-ray image. Its application requires only a well-collimated, low-divergency X-ray beam and an imaging detector with a good spatial resolution like the ones available at PANTER. The intra-focus image is to be recorded at a quite short distance, to be optimized from case to case, from the mirror under test, and the surface roughness must have a negligible effect at the X-ray energy in use. For this reason, the method can be useful for a sensitive mirror shape diagnostics under X-rays, without the need to remove the mirror from the vacuum chamber to re-measure its profile. For this reason, this approach can be very useful in active X-ray optics applications\cite{actop} to derive the mirror shape directly under X-rays, and consequently provide a feedback to the actuator matrix and optimize the mirror shape in real time.

\begin{figure}[H]
	\centering
	\begin{tabular}{c}
     	 \includegraphics[height = 0.4 \textwidth]{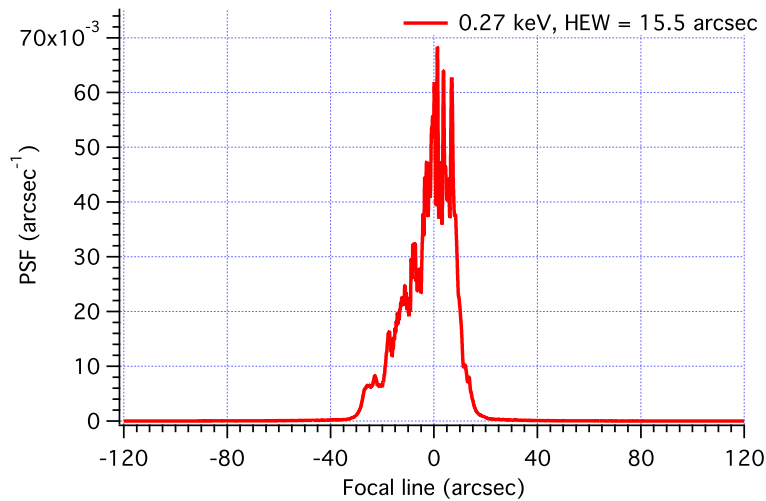}
	\end{tabular}
	\caption{The predicted PSF at 0.27~keV, as computed\cite{RaiSpi2} from the measured microroughness and the longitudinal profiles reconstructed from the intra-focus images.}
	\label{fig:HEW}
\end{figure}

The final purpose of the mirror profile reconstruction is, of course, the imaging quality assessment in focus. Since we have the PP0 mirrors error maps (Fig.~\ref{fig:3D}) compute the double-reflection Point Spread Function (PSF) in focus at 0.27 keV, including the measured roughness PSD\cite{RaiSpi2} even if its impact turns out to be almost negligible at this energy. The resulting PSF is shown in Fig.~\ref{fig:HEW}: it does not include the -- expectedly minor -- effects of roundness errors. {\it The HEW (Half-Energy-Width) computed from this PSF is 15.5~arcsec at 0.27 keV}. This is the value of the angular resolution (Tab.~\ref{tab:XOUs}) expected from the profile reconstruction of the PP0 of the PoC\#2, moreover very close to the HEW value inferred from mirror surface direct mapping\cite{Civitani} (HEW $\approx$ 17~arcsec) and from UV measurements in focus\cite{Civitani} (HEW$\approx$ 16~arcsec at 365 nm, after subtracting in quadrature the aperture diffraction term). This result shows a very important step forward in the angular resolution achieved with X-ray optics manufactured with hot slumped glasses. In fact, this angular resolution is very close to the one of replicated optics in Nickel like those of XMM-Newton. Direct X-ray tests, to be performed in focus at extended PANTER\cite{PANTERext}, are foreseen for January 2014 and are expected to provide the final confirmation that the predictions are correct.
\vspace{-5mm}

The same analysis cannot be performed on the other intra-focal images of the inner layers of the PoC\#2, because in single-reflection setup more than 80\% of the beam would be obstructed. This is a consequence of the dense mirror stacking, which is designed to optimize the on-axis effective area and, at the same time, to enable the desired field of view by keeping unobstructed the off-axis, doubly-reflected rays. It is not, however, optimized to leave unobstructed the singly-reflected beams. Future work will be aimed at extending the analysis of the intra-focal traces to include the computation of the azimuthal errors, and to devise out a way to extend the method to the inner plate pairs in a stacked optical module.

\acknowledgments
This work is supported by the ESA contract No. 22545.

\bibliographystyle{spiebib}

\end{document}